\documentclass[10pt,conference]{IEEEtran}
\usepackage{cite}
\usepackage{amsmath,amssymb,amsfonts}
\usepackage[hyphens]{url}
\usepackage{fancyhdr}
\usepackage{hyperref}

\usepackage{comment}
\usepackage{algorithmic}
\usepackage{graphicx}
\usepackage{caption}
\usepackage{textcomp}
\usepackage{xcolor}
\usepackage[normalem]{ulem}
\usepackage{bbding}
\usepackage{enumitem}
\usepackage{pifont}
\usepackage[linesnumbered, ruled, vlined]{algorithm2e}
\usepackage[makeroom]{cancel}
\usepackage{extarrows}
\usepackage{multirow}
\usepackage{booktabs}
\usepackage{tabularx}
\usepackage[hang, tight]{subfigure}
\usepackage{listings}
\usepackage{xcolor}
\usepackage{color}
\usepackage{colortbl}
\usepackage{tikz}
\usepackage{xcolor}

\newcommand{\revise}[1]{\textcolor{black}{#1}}

\newcommand{\reviseA}[1]{\textcolor{black}{#1}}

\newcommand{\reviseB}[1]{\textcolor{black}{#1}}

\newcommand{\reviseC}[1]{\textcolor{black}{#1}}

\newcommand{\reviseD}[1]{\textcolor{black}{#1}}

\newcommand{\reviseE}[1]{\textcolor{black}{#1}}

\newcommand*\circled[1]{\tikz[baseline=(char.base)]{
            \node[shape=circle,fill,inner sep=1pt,scale=0.8] (char) {\textcolor{white}{#1}};}}

\newcommand*\nofillcircled[1]{\tikz[baseline=(char.base)]{
            \node[shape=circle, draw=black, line width=0.5pt, inner sep=1pt, scale=0.8] (char) {\textcolor{black}{#1}};}}

\usepackage{soul}
\usepackage{siunitx}
\usepackage{multicol}

\newcommand{\hpcayear}{2026}

\newcommand{\hpcasubmissionnumber}{847}
\title{Towards Compute-Aware In-Switch Computing for LLMs Tensor-Parallelism on Multi-GPU Systems}
\usepackage{bbding}

\def\hpcacameraready{} % Uncomment to build camera-ready version

\newcommand\hpcaauthors{Chen Zhang$^1$, Qijun Zhang$^{1}$\textsuperscript{*}\textsuperscript{$\dagger$}, Zhuoshan Zhou$^1$, Yijia Diao$^1$, Haibo Wang$^2$, Zhe Zhou$^2$, Zhipeng Tu$^2$, \\Zhiyao Li$^2$, Guangyu Sun$^3$, Zhuoran Song$^1$, Zhigang Ji$^1$, Jingwen Leng$^1$\textsuperscript{*}, Minyi Guo$^1$}
\newcommand\hpcaaffiliation{Shanghai Jiao Tong University$^1$, Huawei Technologies Co. Ltd.$^2$, Peking University$^3$}
\newcommand\hpcaemail{\{chenzhang.sjtu, zs.zhou, diao\_yijia, songzhuoran, zhigangji\}@sjtu.edu.cn, qijunzhang2000@gmail.com, \\ gsun@pku.edu.cn, \{wanghaibo33, zhouzhe22, tuzhipeng3, lizhiyao5\}@huawei.com, \{leng-jw, guo-my\}@cs.sjtu.edu.cn}

\author{
  \ifdefined\hpcacameraready
    \IEEEauthorblockN{\hpcaauthors{}}
      \IEEEauthorblockA{
        \hpcaaffiliation{} \\
        \hpcaemail{}
      }
  \else
    \IEEEauthorblockN{\normalsize{HPCA \hpcayear{} Submission
      \textbf{\#\hpcasubmissionnumber{}}} \\
      \IEEEauthorblockA{
        Confidential Draft \\
        Do NOT Distribute!!
      }
    }
  \fi 
}

\fancypagestyle{camerareadyfirstpage}{%
  \fancyhead{}
  
  \fancyhead[C]{
    \ifdefined\aeopen
    \parbox[][12mm][t]{13.5cm}{\hpcayear{} IEEE International Symposium on High-Performance Computer Architecture (HPCA)}    
    \else
      \ifdefined\aereviewed
      \parbox[][12mm][t]{13.5cm}{\hpcayear{} IEEE International Symposium on High-Performance Computer Architecture (HPCA)}
      \else
      \ifdefined\aereproduced
      \parbox[][12mm][t]{13.5cm}{\hpcayear{} IEEE International Symposium on High-Performance Computer Architecture (HPCA)}
      \else
      \parbox[][0mm][t]{13.5cm}{\hpcayear{} IEEE International Symposium on High-Performance Computer Architecture (HPCA)}
    \fi 
    \fi 
    \fi 
    \ifdefined\aeopen 
      \includegraphics[width=12mm,height=12mm]{ae-badges/open-research-objects.pdf}
    \fi 
    \ifdefined\aereviewed
      \includegraphics[width=12mm,height=12mm]{ae-badges/research-objects-reviewed.pdf}
    \fi 
    \ifdefined\aereproduced
      \includegraphics[width=12mm,height=12mm]{ae-badges/results-reproduced.pdf}
    \fi
  }
  \fancyfoot[C]{}
}
\begin{document}

\newpage
\maketitle
\setcounter{page}{1}

%Enables the camera ready header and footer
\ifdefined\hpcacameraready 
  \thispagestyle{camerareadyfirstpage}
  \pagestyle{empty}
\else
  \thispagestyle{plain}
  \pagestyle{plain}
\fi

\newcommand{\hpcaheight}{0mm}
\ifdefined\eaopen
\renewcommand{\hpcaheight}{12mm}
\fi

%%%%%% -- PAPER CONTENT STARTS-- %%%%%%%%

\begin{abstract}

Tensor parallelism (TP) in large-scale LLM inference and training introduces frequent collective operations that dominate inter-GPU communication. While in-switch computing, exemplified by NVLink SHARP (NVLS), accelerates collective operations by reducing redundant data transfer, its communication-centric design philosophy introduces the mismatch between its communication mode and the memory semantic requirement of LLM's computation kernel. Such a mismatch isolates the compute and communication phases, resulting in underutilized resources and limited overlap in multi-GPU systems. 

To address the limitation, we propose CAIS, the first \underline{C}ompute-\underline{A}ware \underline{I}n-\underline{S}witch computing framework that aligns communication modes with computation's memory semantics requirement. CAIS consists of three integral techniques: (1) compute-aware ISA and microarchitecture extension to enable compute-aware in-switch computing. (2) merging-aware TB \reviseB{(Thread Block)} coordination to improve the temporal alignment for efficient request merging. (3) graph-level dataflow optimizer to achieve a tight cross-kernel overlap. Evaluations on LLM workloads show that CAIS achieves 1.38× average end-to-end training speedup over the SOTA NVLS-enabled solution, and 1.61× over T3, the SOTA compute-communicate overlap solutions but do not leverage NVLS, demonstrating its effectiveness in accelerating TP on multi-GPU systems.

\end{abstract}

\begingroup\renewcommand\thefootnote{}
\footnotetext{* ~Corresponding author.}
\endgroup

\begingroup\renewcommand\thefootnote{}
\footnotetext{$\dagger$ ~Qijun Zhang led this project during his internship at Shanghai Jiao Tong University.}
\endgroup

\section{Introduction}
\label{section1}

\reviseD{The rapid scaling of large language models (LLMs) has pushed distributed training and inference systems to unprecedented scales, where clusters composed of hundreds or even thousands of GPUs must work together seamlessly~\cite{vaswani2017attention,dubey2024llama,liu2021swin,dosovitskiy2020image}. To fully utilize such large-scale GPU clusters, hybrid parallelism, combining data parallelism (DP), pipeline parallelism (PP), and tensor parallelism (TP), has become the de facto strategy for scaling. Among them, DP and PP mainly serve to scale out across nodes by distributing data batches and network layers, whereas TP is designed to scale up by partitioning large matrix operations across multiple GPUs~\cite{shoeybi2019megatron,narayanan2021efficient,korthikanti2023reducing,liao2025ub}. This tensor-level partitioning enables fine-grained parallel execution but also makes TP the most communication-intensive and structurally complex scheme. A recent study~\cite{liao2025ub} reveals that TP contributes to over 99\% of total data traffic, while data and pipeline parallelism together account for less than 1\%. Furthermore, TP exhibits structurally complex dependencies, as its communication lies on the critical path where computation cannot directly overlap, making optimization inherently difficult~\cite{pati2024t3}. Therefore, as models grow to trillion-parameter scale, TP’s communication overhead becomes a first-order bottleneck that fundamentally limits cluster scalability and system efficiency~\cite{jangda2022breaking,pati2024t3}.}

\reviseD{To overcome this limitation, recent GPU architectures have evolved toward tighter interconnect integration. High-bandwidth NVLink and NVSwitch~\cite{nvl72,dtx} have become essential enablers for scaling up LLM systems through tensor parallelism, providing terabyte-per-second inter-GPU connectivity.
However, as LLMs continue to grow, even these high-speed links face severe pressure from redundant data transfers among GPUs. To further relieve this bottleneck, NVIDIA introduced NVLink SHARP (NVLS)~\cite{nvls,klenk2020network}, an in-switch computing architecture that performs collective reductions directly within the NVSwitch fabric.}
%To mitigate the communication overheads, current multi-GPU systems introduce in-switch computing techniques, exemplified by NVIDIA's NVLink SHARP~\cite{nvls,klenk2020network} (NVLS), offloading collective operations (e.g., AllReduce, Reduce-Scatter) to the NVSwitch fabric. 
By performing reductions in-flight, NVLS reduces redundant GPU-to-GPU transfers and achieves 2–8× speedups for collective primitives compared to GPU-driven implementations~\cite{klenk2020network}. However, NVLS remains fundamentally communication-centric, focusing solely on accelerating collective operations without considering their interplay with computation kernels such as GEMM. This isolation prevents smooth and tight overlap of communication and computation, often leaving GPUs idle during communication phases. This problem is further amplified in large-scale LLM inference and training workloads, where tensor parallelism introduces frequent collective operations that account for up to 40–60\% of total latency~\cite{pati2024t3,jangda2022breaking,wang2022overlap}. 
While recent work has explored compute-communication overlap through software scheduling~\cite{jangda2022breaking,wang2022overlap,chen2024centauri} and hardware-assisted overlapping~\cite{pati2024t3}, none leverage in-switch computing, missing the opportunity to fully exploit NVLS’s architectural potential.

The key obstacle lies in the mismatches between the existing NVLS communication primitives and the requirements of TP compute kernels: \emph{the communication mode (i.e., push/pull modes) of NVLS primitives fail to align with the required memory semantics (i.e., read/write) of TP kernels such as GEMM.} 
% Although NVLS supports both push and pull modes (e.g., \texttt{multimem.st} and \texttt{multimem.ld\_reduce}), its memory semantics fail to align with the remote read/write behavior of TP kernels such as GEMM. 
For example, an AllGather operation followed by a GEMM often requires on-demand remote \textit{reads}, but NVLS implements AllGather as \textit{push}-based stores with the \texttt{multimem.st} instruction, transmitting data eagerly regardless of when the consumer computation is ready. Likewise, a GEMM followed by Reduce-Scatter requires distributed \textit{writes}, but NVLS implements it with a \textit{pull}-based \texttt{multimem.ld\_reduce} instruction, which forces consumers to fetch data instead of receiving it inline. This misalignment introduces strict global barriers that prevent fine-grained compute-communication overlap. 

This limitation motivates a shift toward compute-aware in-switch computing that aligns communication modes with computation’s memory semantics requirement. Therefore, we propose CAIS, the first \underline{C}ompute-\underline{A}ware \underline{I}n-\underline{S}witch computing framework that aligns the two sides. %communication modes with computation’s memory semantics requirement. 
With CAIS, the computation kernel should directly issue load/reduction instructions for communication following its memory semantic requirement, while the switch automatically performs request merging for these remote accesses. 

However, designing such a compute-aware in-switch computing has three challenges: \nofillcircled{1} Current GPU and NVLS lack necessary ISA and architecture support to express and process compute-aware in-switch computing. \nofillcircled{2} Even with instruction and architecture supports, independently scheduled TBs across GPUs result in staggered requests, reducing merge efficiency and causing switch buffer contention. \nofillcircled{3} Isolated operators make it difficult to exploit the producer-consumer relationships in LLM dataflow graph (DFG), limiting resource utilization. To address these challenges, CAIS co-designs GPU ISA, switch microarchitecture, and graph-level dataflow with three techniques: \circled{1} CAIS provides compute-aware ISA and microarchitecture extension that enables compute-aware in-switch computing. \circled{2} CAIS introduces a lightweight compiler-architecture co-design to coordinate TB execution across GPUs, maximizing merge success rate without incurring high synchronization costs. \circled{3} CAIS integrates a graph-level dataflow optimizer, exploiting fine-grained TB-level overlapped execution to maximize computation and communication resource utilization. 
To the best of our knowledge, this is the first work to realize fine-grained compute-communication overlap within an in-switch computing paradigm. %\looseness=-1

We summarize the key contributions as follows:
\begin{itemize}
    \item We uncover a critical misalignment between the communication semantics of current in-switch primitives and the memory access requirements of LLM compute kernels, resulting in underutilized resources and limited overlap. 
    \item We propose CAIS, the first compute-aware in-switch computing framework that co-designs GPU ISA, switch microarchitecture, and operator dataflows to enable fine-grained, kernel-integrated compute-communicate overlap. 
    \item We implement CAIS in a cycle-accurate simulator and evaluate it on three LLM inference and training workloads, achieving on average 1.38× end-to-end training speedup over NVLS-augmented baselines, and 1.61× over T3~\cite{pati2024t3}, the SOTA compute-communicate overlap solutions but do not leverage NVLS. 
\end{itemize}

The rest of this paper is organized as follows. Section~II reviews background and motivates compute-aware in-switch computing. Section~III presents the CAIS design, including ISA extensions, micro-architecture design, and the supporting compiler and runtime mechanisms. Section~IV describes our experimental methodology. Section~V evaluates CAIS on representative LLM training and inference workloads. Section~VI discusses related work and Section~VII concludes.

\section{Background and Motivation}

\begin{figure*}[!t]
\includegraphics[width=1\textwidth]{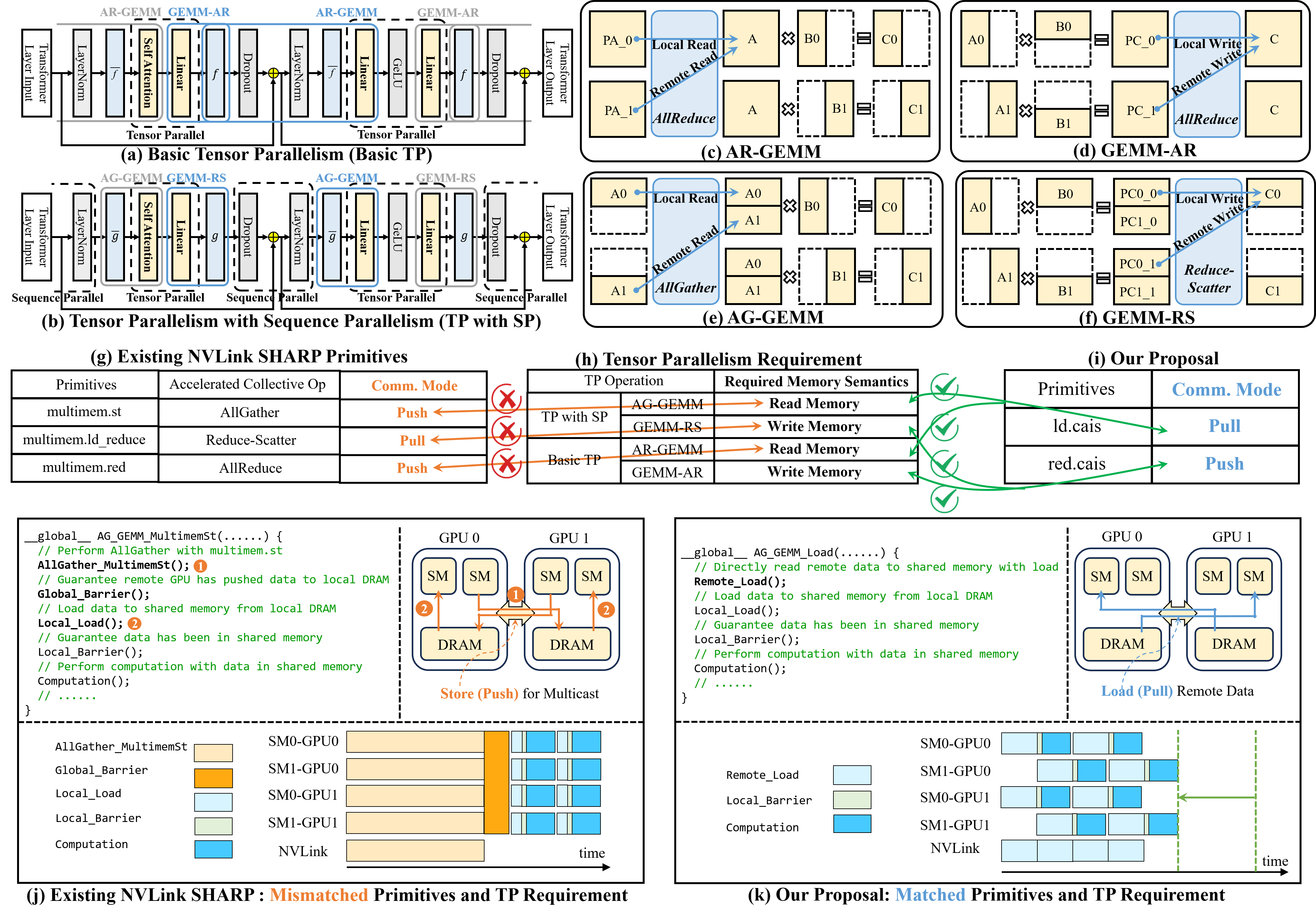}
%\vspace{-.1in}
\caption{Motivation for Compute-Aware In-Switch Computing in Tensor Parallelism. (a–b) Tensor Parallelism (TP) in LLM. (c–f) Collective Communication and Computation Kernel Relationship. (g–i) Comparison of Existing NVLS Primitives, Compute-aware TP Requirements and Our Proposal.
(j-k) Comparison of Computation Details between Existing NVLS and Our Proposal. }
\label{motivation}
%\vspace{-.25in}
\end{figure*}

\subsection{LLM and Tensor Parallelism}

Large language models (LLMs) such as GPT-4~\cite{achiam2023gpt} and LLaMA-3~\cite{grattafiori2024llama} have surpassed the trillion-parameter scale, driving unprecedented demands for compute and memory bandwidth. These models are dominated by dense linear algebra kernels, particularly general matrix multiplications (GEMMs) and matrix-vector multiplications (GEMVs), which exhibit quadratic or cubic growth in compute and memory complexity as hidden dimensions or attention heads increase. To scale these workloads across multiple GPUs, modern deployments employ hybrid parallelism strategies combining data parallelism (DP)~\cite{rajbhandari2020zero,rasley2020deepspeed}, pipeline parallelism (PP)~\cite{huang2019gpipe,li2021chimera}, and tensor parallelism (TP)~\cite{shoeybi2019megatron,narayanan2021efficient,korthikanti2023reducing}.

Among these strategies, TP introduces the heaviest communication overhead, because each FFN or attention layer is split across GPUs and requires frequent inter-GPU collective operations, e.g., AllReduce and AllGather, to aggregate partial results. Unlike DP, which incurs communication only during gradient synchronization, or PP, where communication is confined to activation exchanges between stages, TP creates per-layer synchronization points that scale with model depth and width. Recent studies show that for multi-GPU training with TP, 40–60\% of end-to-end latency arises from inter-GPU data transfers~\cite{pati2024t3,jangda2022breaking,wang2022overlap}. As LLMs scale, communication costs are expected to dominate overall runtime unless new architectural solutions emerge.

Fig.~\ref{motivation}(a)(b) illustrates two commonly-used TP strategies.
(1) Basic TP partitions the key computational modules of a Transformer, such as the Q, K, and V in Attention, as well as the FFN, across multiple GPUs along the hidden dimension or the head dimension. In Fig.~\ref{motivation}(a), $f$ is AllReduce (AR) in the forward pass and no operation in the backward pass. $\overline{f}$ is no operation in the forward pass and AllReduce in the backward pass. 
(2) TP with Sequence Parallelism (SP), a more recent variant, partitions operations along the sequence length and employs a combination of AllGather (AG) and Reduce-Scatter (RS) operations to eliminate redundant activations. In Fig.~\ref{motivation}(b), $g$ is Reduce-Scatter
in the forward pass and AllGather in the backward pass. $\overline{g}$ is AllGather in the forward pass and Reduce-Scatter in the backward pass. Although AllReduce in Basic TP is mathematically equivalent to Reduce-Scatter plus AllGather, TP with SP can partition more operations (e.g., LayerNorm) and hence reduces memory consumption for activations across GPUs.

\subsection{NVLink/NVSwitch-based Multi-GPU Systems}
\label{section2B}

\begin{figure}[!t]
\includegraphics[width=0.5\textwidth]{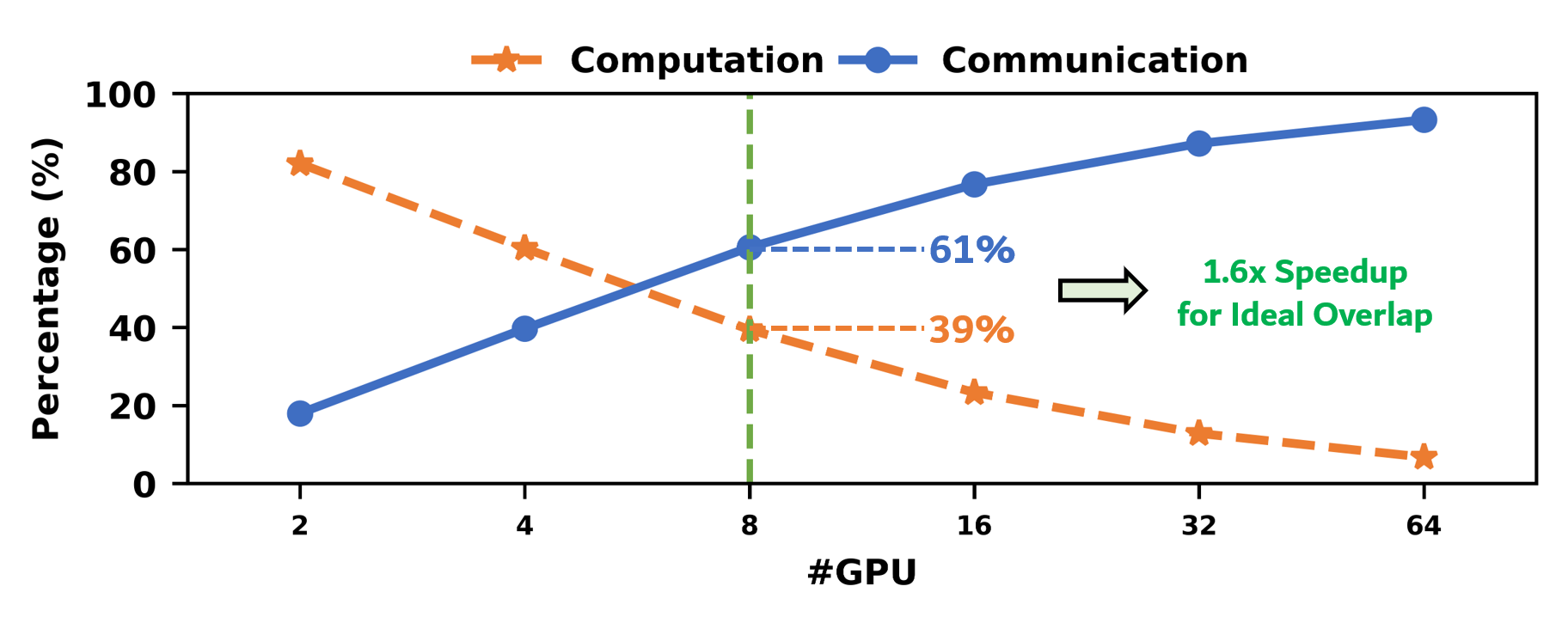}
%\vspace{-.3in}
\caption{\reviseB{Computation-Communication Time When Scaling Up.}}
\label{compcomm}
%\vspace{-.25in}
\end{figure}

Modern AI systems attempt to address communication bottlenecks by coupling dozens or hundreds of GPUs via high-radix NVLink/NVSwitch networks~\cite{nvl72,dtx}. NVLink has evolved from its first generation~\cite{foley2017ultra}, delivering 160 GB/s GPU-to-GPU bandwidth on Pascal, to the fifth generation~\cite{nvl72} in Blackwell, which provides 1.8 TB/s GPU-to-GPU bandwidth and powers large-scale systems such as NVL72 (72 GPUs)~\cite{nvl72,dtx}. 
While these fabrics offer scalable collective operations, their performance remains bounded by link bandwidth. 
\reviseE{To quantify this limitation, we execute LLaMA-7B on our simulated NVIDIA H100 SuperPOD interconnected via a 900 GB/s NVLink/NVSwitch fabric, varying the number of participating GPUs (see Section \ref{sec:methodology}).} \reviseB{As shown in Fig.~\ref{compcomm}, communication time quickly overtakes computation time once the system scales beyond 4–8 GPUs; In particular, under an 8-GPU configuration, the average communication time is about 1.6× longer than computation across the model.} 
This problem will worsen with future 1T+ parameter models, whose communication volume grows super-linearly due to deeper layers and larger token batches. These observations underscore the urgent need for architectural approaches that reduce or hide communication, rather than merely speeding up links.

In-switch computing has attracted much attention in the computer network community. Many works~\cite{graham2016scalable,liu2017incbricks,jin2017netcache,li2019accelerating,gebara2021network,lao2021atp,sapio2021scaling,de2021flare,fei2021efficient,yuan2022unlocking,yang2022using,he2023generic,liu2023network,haghi2023flash,wang2023roar,haghi2024smartfuse} have been proposed to accelerate the AllReduce in the distributed system. In recent years, NVIDIA's NVLink SHARP~\cite{nvls,klenk2020network} (NVLS) brought in-switch computing into inter-chip network to address these efficiency and scalability bottlenecks of multi-GPU systems. 
NVLS offloads collective operations (e.g., AllReduce and Reduce-Scatter) to NVSwitch, performing reductions ``in-flight'' and reducing data movement~\cite{klenk2020network,nvls}. NVLS has been supported in modern GPU architecture. With NVIDIA’s Hopper GPUs, in-switch operations such as multicast and reduction can be issued via PTX-level \texttt{multimem} instructions, including \texttt{multimem.st}, \texttt{multimem.ld\_reduce}, and \texttt{multimem.red}, enabling collective operations to be performed inside NVSwitch fabrics. These instructions can, in principle, be embedded in computation kernels such as GEMM to trigger multi-GPU collectives directly, and have become a cornerstone capability in modern systems. The study~\cite{klenk2020network} on NVLS has demonstrated 2×–8× speedups for collective operations compared to GPU-driven communication, thanks to its hardware-accelerated multicast and reduction integrated with NVSwitch.

\subsection{Limits of Current In-Switch Computing}

Despite these advances, current in-switch computing remains fundamentally communication-centric: it is solely designed to accelerate collective operations but remains agnostic to the computation kernels such as GEMM that produce or consume these data streams. For example, AllGather operator collects data chunks from all participating GPUs and redistributes the complete concatenated result back to every GPU. NVLS currently implements this via the \texttt{multimem.st} instruction, where each GPU proactively “pushes” its data to all other GPUs as soon as it becomes available. In the context of pure collective communication acceleration for the AllGather operator, push mode provides clear advantages over pull mode. 
By allowing each GPU to proactively transmit its data to peers as soon as it is ready, push mode forms a continuous, one-way data stream that avoids the roundtrip latency to remote memory inherent in pull-based communication. This also ensures higher bandwidth utilization, as the data pipeline remains fully saturated without idle periods caused by prolonged traffics. 
Other collective operations follow a similar design paradigm. Fig.~\ref{motivation}(g)'s table lists all the NVLS PTX primitives and their corresponding communication modes.

However, this communication-centric design \textbf{limits the opportunities for overlapping computation and communication}. For instance, during the forward pass of LLMs, AllGather operation is often immediately followed by a GEMM operator, such as in attention and FFN. As illustrated in Fig.~\ref{motivation}(e), the GEMM's computation requires memory \textbf{reads} from both local and remote devices. Yet, \texttt{multimem.st}, the in-switch computing instruction used for AllGather, operates in \textbf{push-mode}. 
This mismatch between architectural support and workload requirements forces computation and its associated communication to be executed on different GPUs. It introduces global barriers to preserve producer-consumer dependencies.
As a result, computation and communication are isolated into separate phases, leaving SMs idle while waiting for synchronization. Fig.~\ref{motivation}(j) visualizes this process in detail. Profiling on LLaMA-7B with TP shows that GPU utilization can drop below 60\%, even when NVLS is enabled.

To summarize the scope of this mismatch, Fig.~\ref{motivation}(c)–(f) illustrate the memory access patterns of the four computation-communication combinations in TP across multiple GPUs, while Fig.~\ref{motivation}(g)–(h) summarize the mismatches between the current NVLS design and the requirements of computation kernels. Specifically, AllGather + GEMM (AG-GEMM) requires memory reads, but NVLS only provides \texttt{multimem.st} in push mode. GEMM + Reduce-Scatter (GEMM-RS) requires memory writes, but NVLS provides only \texttt{multimem.ld\_reduce} in pull mode. Similarly, Basic TP with AllReduce + GEMM (AR-GEMM) and GEMM + AllReduce (GEMM-AR) requires both memory reads and writes, while NVLS currently offers only \texttt{multimem.red} in push mode.

\subsection{Design Philosophy and Challenges}
\label{section2D}

\begin{figure}[!t]
\centering
\includegraphics[width=0.5\textwidth]{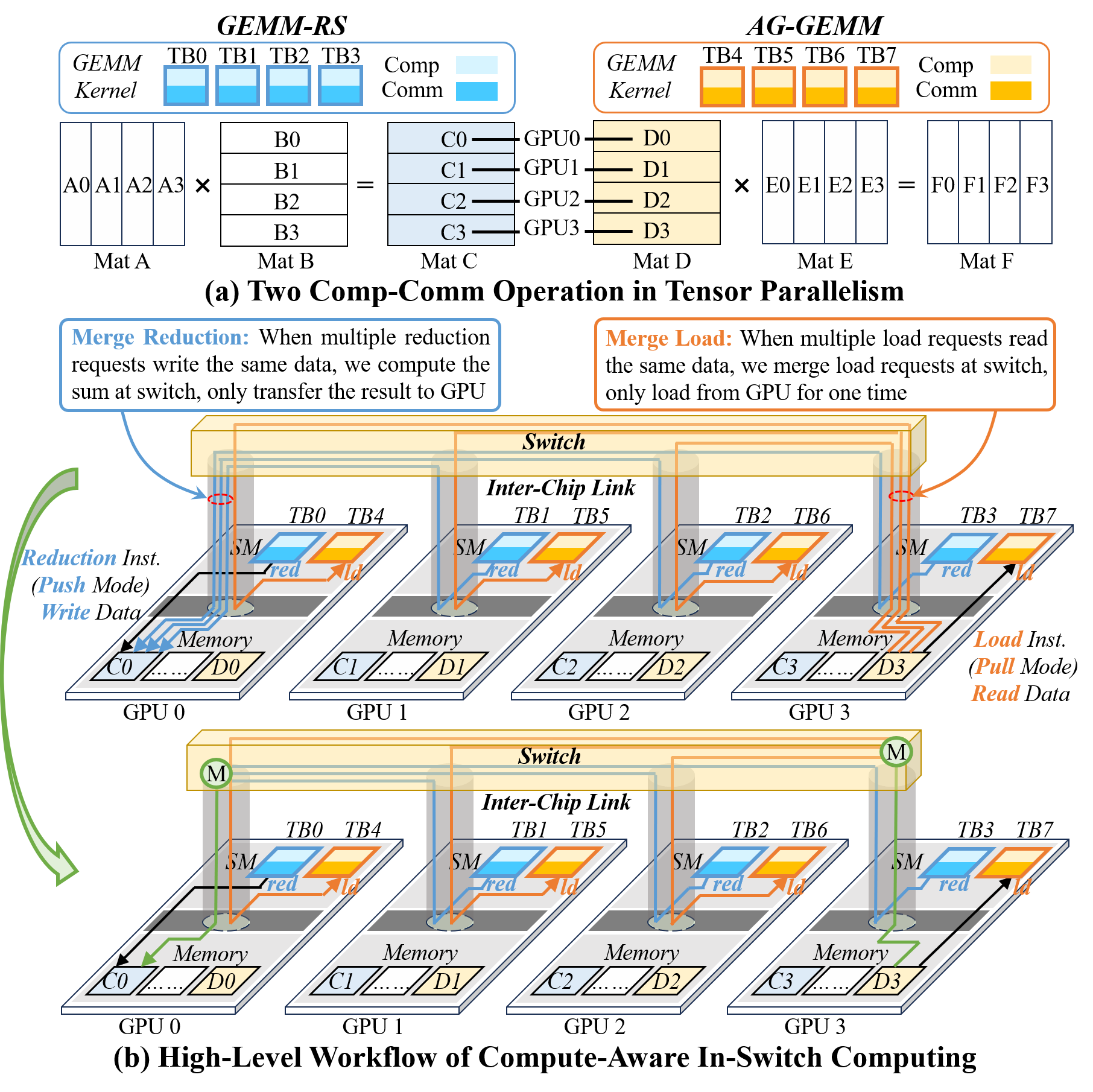}
%\vspace{-.12in}
\caption{The System Architecture of CAIS.}
\label{principle}
%\vspace{-.25in}
\end{figure}

As previously analyzed, a fundamental mismatch exists between the communication modes supported by current in-switch computing systems and the memory access semantics required by LLM computational kernels, resulting in isolated computation and communication. This limitation motivates a shift toward \emph{compute-aware in-switch computing}. Our philosophy is: 
\textit{computation kernel should directly issue load/reduction instructions for communication following its memory semantic requirement, while the switch automatically performs request merging for these remote accesses.} 

This approach enables semantic alignment between communication mode and computational intent, unlocking more native and tighter communication-computation overlap. As illustrated in Fig.~\ref{motivation}(k), taking AG-GEMM as an example, following its memory semantic requirement, computation kernel directly reads remote data via load operations in pull mode. Because both the computation and its remote data access are performed by the same TB, only TB-level local barriers are needed to enforce dependencies between communication and computation, allowing computation and communication across different SMs to overlap naturally.

%Fig.~\ref{principle} illustrates the architectural functionality of compute-aware in-switch computing. 
The system architecture of CAIS is illustrated in Fig.~\ref{principle}. When a GEMM kernel issues reduction or load instructions (for GEMM-RS and AG-GEMM operations, respectively) to access remote data, the switch dynamically merges these requests to reduce communication overhead: (1) Reduction requests: When multiple reduction requests target the same data, the switch aggregates them by computing the sum and transmitting only the final result to the destination GPU, thus reducing the downstream traffic from switch to GPU. 
(2) Load requests: When multiple load requests reference the same data, the switch fetches the data only once from the target GPU and replicates it to the requesting GPUs, reducing the upstream traffic from GPU to switch.

However, designing such a compute-aware in-switch computing has three architectural challenges: 

\textbf{(1) Lack of ISA and Microarchitecture Support.} 
Without GPU ISA and switch microarchitecture support, commercial GPU cannot perform compute-aware in-switch computing. To address this limitation, we propose PTX-level instruction extensions for compute-aware in-switch computing and provide microarchitectural support for request merging by integrating a hardware merge unit into the data path at the switch port. These two supports enable the functionality of the compute-aware in-switch computing.

\textbf{(2) Lack of Temporal Coordination.} Even with semantic alignment, temporal misalignment across GPUs~\cite{jain2024pal} remains a critical bottleneck. Thread blocks (TBs) on different GPUs are scheduled independently, resulting in staggered memory requests to the same remote addresses. This misalignment prevents the switch from merging requests effectively, as early-arriving requests must wait for delayed ones, leading to buffer contention and eviction. \reviseB{Our simulation shows that the delay between the earliest and latest requests to the same address averages 35 µs.} We address this by introducing a compiler-hardware co-design strategy. The compiler statically groups TBs with shared data dependencies, while the GPU runtime enforces lightweight TB-group synchronization, aligning access timing and improving merge success rates. \reviseB{Our experiments demonstrate that these optimizations reduce the delay to 3 µs, achieving about 10× improvement.}

\textbf{(3) Limited Cross-Kernel Fusion.} Prior in-switch computing approaches have struggled to achieve deep cross-kernel fusion due to the aforementioned misalignment, which forces collective kernels operate as isolated phases, making it difficult to exploit the producer-consumer relationships in the LLM dataflow graph (DFG). By resolving it, CAIS enables deeper DFG-level optimizations, allowing TB-level producer-consumer relationships to be established. As soon as one TB of an operator completes, dependent TBs of subsequent (and even further downstream) operators can launch immediately, which greatly improve the computation-communication efficiency. Moreover, we find that this approach also allows CAIS to fuse operators with complementary communication patterns, maximizing overall bandwidth utilization.

\section{CAIS Design}
\label{section3}

Following the above design philosophy, we introduce CAIS, a compute-aware in-switch computing framework to overcome the limitation of existing communication-centric in-switch computing. The framework consists of three primary components: \reviseA{\emph{1) Compute-Aware ISA and Microarchitecture Extensions}. This is the core design of CAIS, which fundamentally eliminates the global computation–communication barrier by aligning communication modes with the semantic requirements of computation.} 

\reviseA{Building upon the architectural foundation that removes the global barrier, CAIS further integrates two optimizations:} 2) \emph{Multi-GPU TB Coordination} that aligns cross-GPU TB execution using compiler-guided grouping and lightweight in-switch synchronization to maximize temporal locality for request merging. 3) \emph{Graph-Level Dataflow Optimizer} that exploits fine-grained dependency to fuse communication-heavy operator sequences, e.g., GEMM-RS + LN + AG-GEMM, into a single execution pipeline, improving bandwidth utilization and end-to-end performance.

\subsection{Compute-Aware ISA and Microarchitecture Extensions}
\label{archsupport}

To support compute-aware in-switch computing, CAIS introduces a co-designed ISA extension and switch microarchitecture that enable dynamic request merging for both load and reduction operations across GPUs. This design transforms the switch from a passive relay into an active compute-aware merging agent, significantly reducing redundant inter-chip traffic and improving execution efficiency for tensor-parallel (TP) workloads.

\subsubsection{ISA Extention for Mergeable Memory Access}

\begin{figure}[!t]
\includegraphics[width=0.5\textwidth]{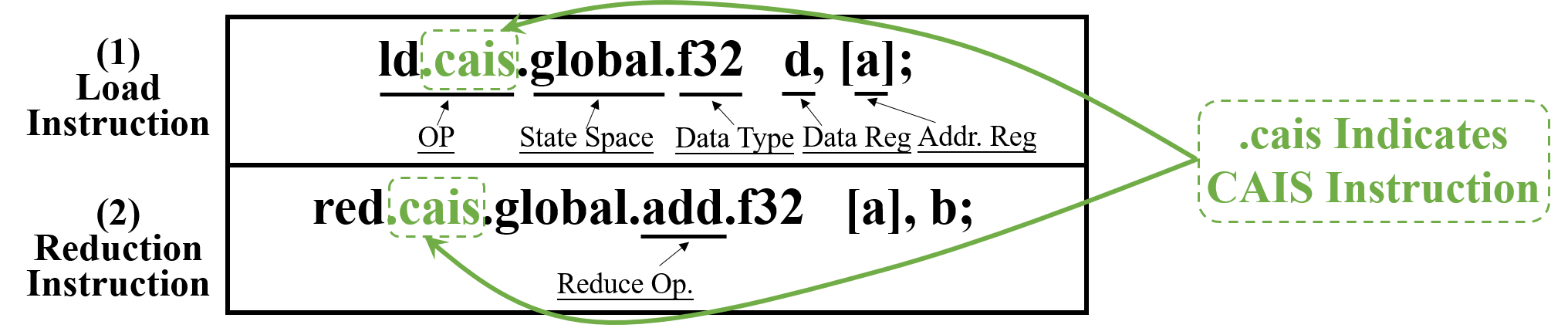}
%\vspace{-.05in}
\caption{Extension of the PTX Instructions.} %We provide CAIS instructions extended from load and reduction instructions for merging.
\label{isa}
%\vspace{-.15in}
\end{figure}

We extend NVIDIA's PTX instruction set with two new instructions: \texttt{ld.cais} and \texttt{red.cais}, as shown in Fig.~\ref{isa}. These instructions encode a 1-bit CAIS flag in memory access requests, signaling the switch that the request is eligible for in-switch merging. This lightweight annotation allows the system to selectively apply merging to communication-intensive operations such as AllGather loads or ReduceScatter reductions, without modifying existing computation semantics.

\subsubsection{Switch Micro-architecture for Request Merging}

\begin{figure}[!t]
\centering
\includegraphics[width=0.5\textwidth]{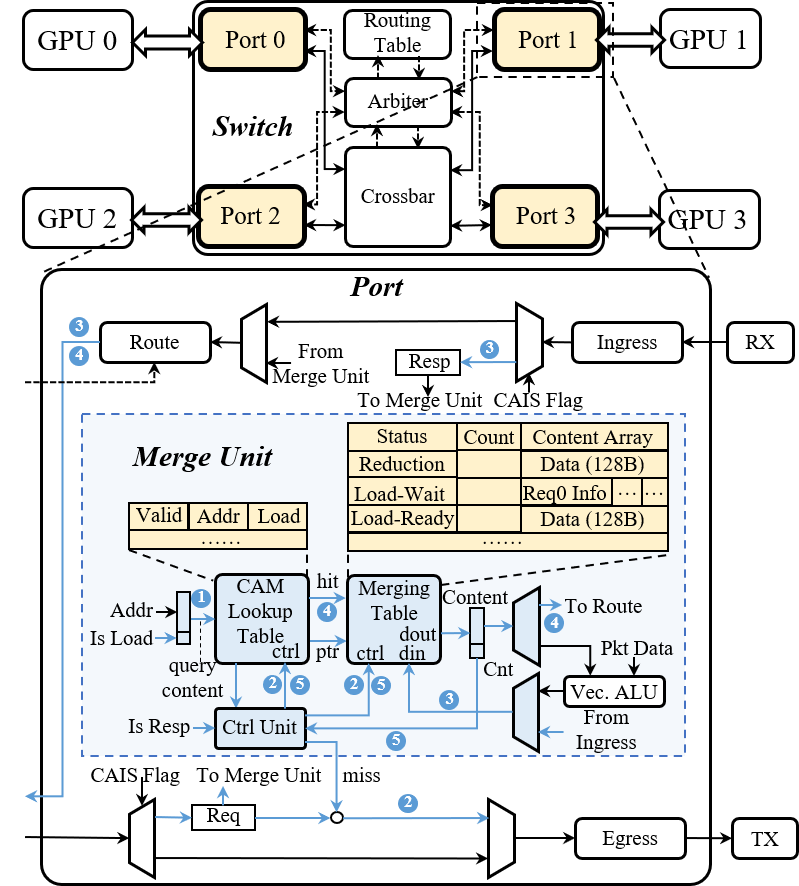}
%\vspace{-.05in}
\caption{Switch Micro-architecture for CAIS.}
\label{switcharch}
%\vspace{-.25in}
\end{figure}

To support CAIS instructions, we enhance NVSwitch datapath with a dedicated merge unit (Fig.~\ref{switcharch}), mainly consisting of two tables: 
1) \textbf{CAM Lookup Table} matches incoming requests based on memory address and type (load or reduction). On a match, request is merged into an existing session; otherwise, a new entry is created. 
2) \textbf{Merging Table} maintains partial results for each session, including cached data for loads or accumulated sums for reductions. Each entry tracks session state (Load-Wait, Load-Ready, or Reduction) and a counter of merged requests. \looseness=-1

These tables operate in tandem to perform on-the-fly aggregation of identical accesses across GPUs. When the last contributing request arrives, the merged data is either forwarded to requesters (loads) or written to memory (reductions).

\subsubsection{In-Switch Micro-Functions for Load and Reduction}
\label{section3A3}

\begin{figure}[!t]
\includegraphics[width=0.48\textwidth]{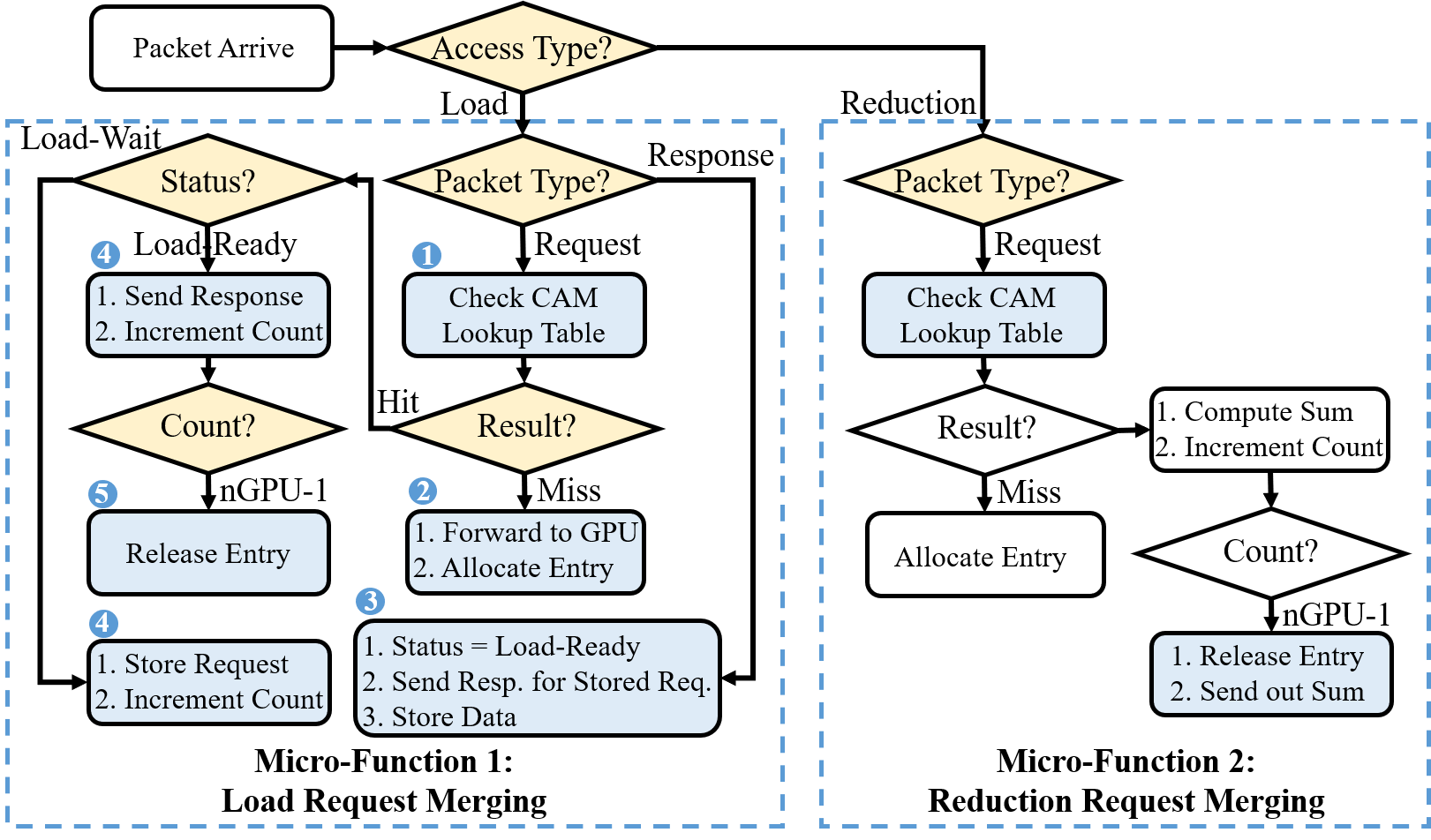}
%\vspace{-.1in}
\caption{In-switch Micro-Functions Workflow.}
\label{microfunction}
%\vspace{-.3in}
\end{figure}

With the ISA and switch microarchitecture extensions, CAIS perform request merging with two micro-functions that handle \textit{load} and \textit{reduction} requests inside the NVSwitch. 
These micro-functions extend the existing NVLS pipeline by performing dynamic request detection, caching, and response generation in-flight, thereby reducing redundant traffic and avoiding unnecessary synchronization. Fig.~\ref{microfunction} illustrates the flow of the two micro-functions.

\reviseB{\textbf{Micro-Function 1: Load Request Merging.} Load request merging eliminates redundant load responses. When a \texttt{ld.cais} request arrives at the switch, the merge unit first performs an associative search within the CAM Lookup Table to \noindent\circled{1} check for an existing merge entry targeting the same memory address and request type. \noindent\circled{2} If no match is found, a new entry is allocated in both the CAM Lookup Table and the Merging Table. The request is forwarded to the destination GPU through the standard routing path, while the new entry in the Merging Table is initialized with ``\texttt{Status} = \texttt{Load-Wait}, \texttt{Count} = 1'', and the associated request metadata is stored in the \texttt{Content Array}. \noindent\circled{3} When the response data from the target GPU returns, the status is updated to \texttt{Load-Ready}, and the data is cached in the \texttt{Content Array}. The switch also generates responses for requests stored in \texttt{Content Array} before caching the arriving data. After that, the switch can serve subsequent requests to the same address directly from this cached data without reissuing memory transactions to the target GPU. \noindent\circled{4} If a later request arrives and hits an active session, the merge unit either appends the request metadata in \texttt{Content Array} for deferred response, if the data is still pending, or otherwise immediately generates a response with the cached data in \texttt{Content Array}. \noindent\circled{5} The completes and its table entries are released once the \texttt{Count} equals the number of participating GPUs minus one, excluding the GPU that holds the local copy.}

\reviseB{\textbf{Micro-Function 2: Reduction Request Merging.} Reduction request merging eliminates redundant reduction requests. Similar to load request merging, for \texttt{red.cais}, multiple contributions to the same address are accumulated directly within the switch. Once all expected requests are received, the sum is written to the destination memory, avoiding duplicate transmissions. The white blocks in Figure~\ref{microfunction} indicate datapaths reused from NVLS.}

Through this combination of load and reduction micro-functions, the switch can dynamically merge multiple remote accesses, turning multiple data transmissions into a single consolidated operation.

\subsubsection{\reviseC{Eviction Mechanism}}
\label{section3A4}

\reviseC{If a new entry must be allocated but the tables are full, an LRU-based eviction policy is triggered. 1) If the selected entry is for reduction merging, it is directly evicted, and the partial result is sent to the home GPU of its address. 2) If the selected entry is for load merging, entries in the \texttt{Load-Ready} state can be safely evicted, whereas those in the \texttt{Load-Wait} state are deferred until the response data arrives. In this case, the arriving pending request bypasses the merge unit without triggering further eviction, avoiding thrashing or deadlock.} 

\reviseC{To handle the remaining requests for the evicted entry, a timeout-based forward-progress mechanism is employed, similar to that in existing NVLS~\cite{klenk2020network}. Each merge entry is equipped with a timer to track the elapsed time since its last access. If this timer exceeds a predefined threshold, the entry is automatically evicted, ensuring that no request remains stalled. } 

\subsubsection{Deterministic Routing for Merging Convergence}

To ensure all mergeable requests targeting the same address converge at the same switch, CAIS adopts a deterministic routing algorithm similar to that used in existing NVSwitch systems~\cite{klenk2020network}. A lightweight hash function on the request address (or a subset of its bits) maps each request to a fixed path, guaranteeing that matching requests are processed by the same merge unit. Since LLM workloads exhibit regular and predictable access patterns, a simple deterministic routing scheme is sufficient to prevent deadlocks and ensure high link utilization without complex path selection.

\subsection{Cross-GPU TB Coordination}
\label{incsync}

While compute-aware ISA and switch microarchitecture provide the foundation for in-switch request merging, their effectiveness critically depends on the temporal alignment of memory requests across GPUs. In the absence of coordination, mergeable load or reduction requests from different GPUs may arrive at the switch at different times, resulting in missed merging opportunities or buffer pressure due to delayed aggregation. This temporal misalignment is rooted in the fact that TBs are independently scheduled by the GPU runtime, leading to execution drift across devices. Even for the same operator, TBs on different GPUs may issue their memory requests at slightly different times. For in-switch merging to be effective, however, these requests must arrive closely in time, within the lifetime of the Merge Table entry. Otherwise, early-arriving requests may be evicted or bypassed before others arrive, negating the benefits of merging. To address this, CAIS introduces a lightweight coordination mechanism at the granularity of thread blocks, the fundamental parallel execution unit in modern GPU workloads. This coordination enforces temporal locality among semantically equivalent memory requests across GPUs, allowing in-switch merging logic to operate with minimal buffering and maximal aggregation efficiency.

\begin{figure}[!t]
\includegraphics[width=0.5\textwidth]{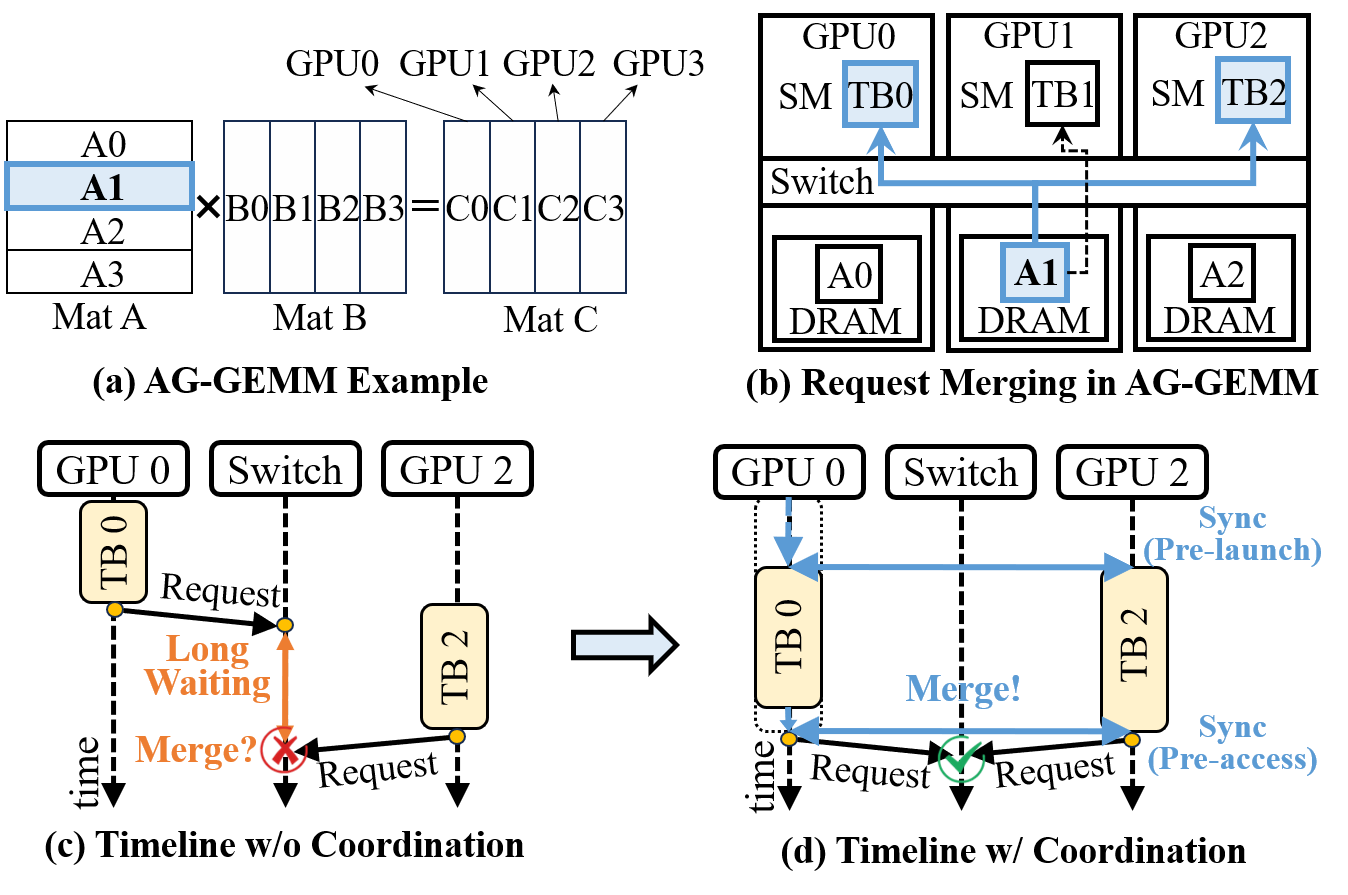}
%\vspace{-.15in}
\caption{Merging-aware TB-Group Coordination.}
\label{tbsync}
%\vspace{-.25in}
\end{figure}

\begin{figure}[!t]
\includegraphics[width=0.5\textwidth]{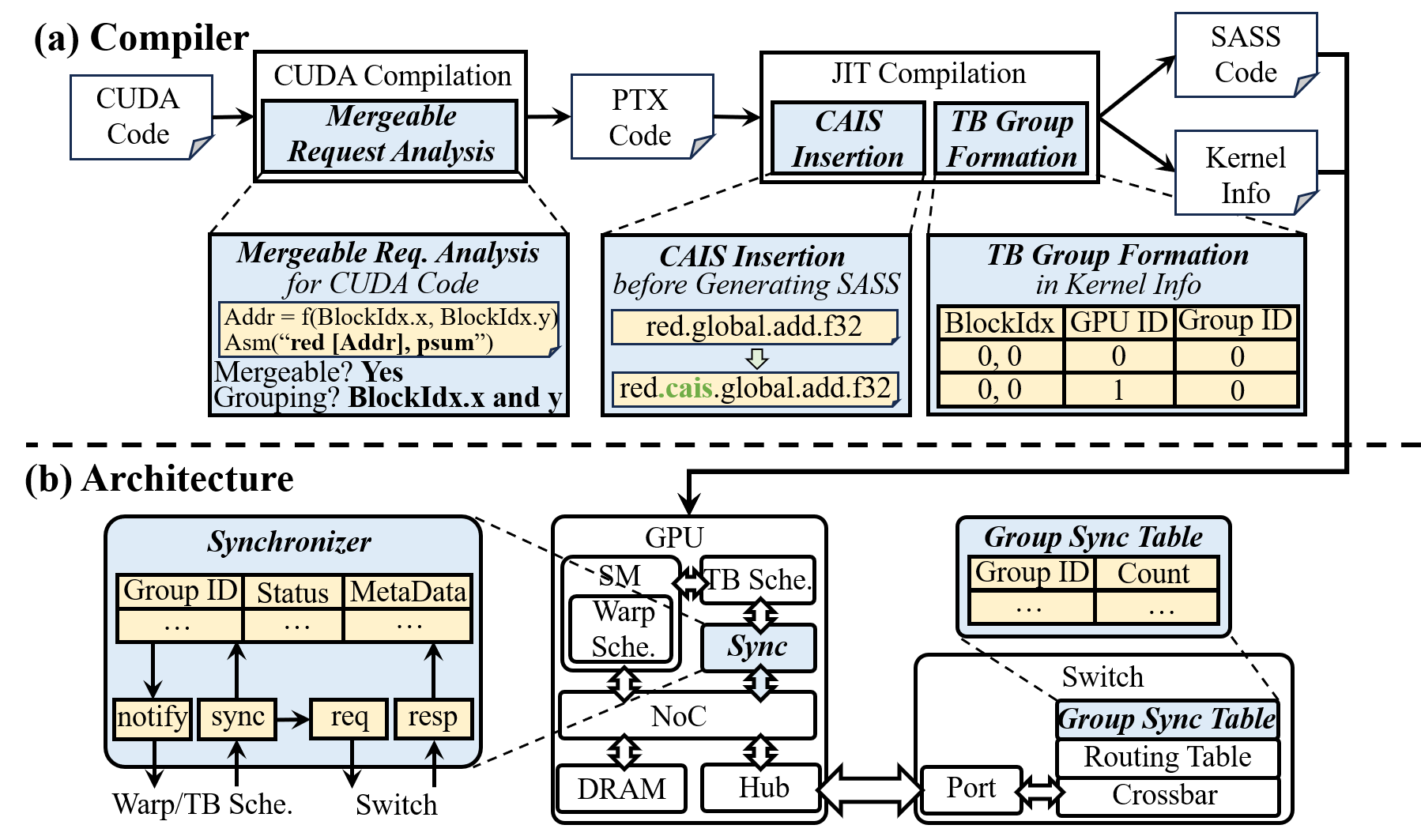}
%\vspace{-.25in}
\caption{Compiler and Architecture Support for TB Coordination. The compiler creates TB Groups according to data dependency, and the architecture uses a synchronizer to align request timing across GPUs.}
\label{tbsyncarch}
%\vspace{-.25in}
\end{figure}

\subsubsection{Compiler-Guided TB Grouping} 
\label{compilersupport}

\textbf{TB-Group Based Coordination.} To ensure temporal locality, CAIS organizes mergeable TBs into logical TB-groups. As is shown in Fig.~\ref{tbsync}(b), each TB-group contains all TBs across GPUs that access the same data region using CAIS-tagged instructions (e.g., \texttt{ld.cais}, \texttt{red.cais}). A switch-side merge tracker monitors request arrival patterns for each group and address. Only when requests from all participating GPUs are observed for a given address does the switch perform merging. On the GPU side, each SM uses lightweight hardware counters to track TB-group progress. Once a TB issues a mergeable request, it waits until a local readiness condition is met, either via a credit mechanism or through switch, issued acknowledgment, ensuring alignment with peer TBs on other GPUs. 

\textbf{Compiler Support.} CAIS leverages compiler assistance to identify TBs that are likely to issue mergeable memory requests and groups them accordingly. As illustrated in Fig.~\ref{tbsyncarch}(a), during the CUDA-to-PTX compilation stage, 
\reviseC{we perform static index analysis on the address expressions of memory access instructions. The analysis detects whether the expression contains the GPU ID. If not, the index is GPU-invariant, indicating that the instruction will access the same memory location when other factors in the expression, such as \texttt{blockIdx}, are identical. Consequently, TBs across different GPUs but with the same \texttt{blockIdx} are expected to access identical data.} 
The compiler then groups such TBs into logical TB Groups. During JIT compilation, memory access instructions associated with these groups are replaced by their CAIS variants (e.g., \texttt{ld.cais}, \texttt{red.cais}). Additionally, the compiler attaches TB Group metadata to the kernel launch configuration, which is used by the runtime and switch to guide synchronization and merging behavior.

\subsubsection{TB Group-Aware Synchronization Mechanisms}
\label{section3B2}

To ensure temporal alignment, CAIS introduces two synchronization mechanisms, as shown in Fig.~\ref{tbsync}(d): 
(1) \textbf{Pre-launch synchronization.} Before a TB is dispatched by the GPU's scheduler, it registers its Group ID with the local synchronizer. This synchronizer then issues a synchronization request to the switch. The TB remains in a \textit{pending} state until the switch confirms that all GPUs in the group have registered corresponding TBs. Once the switch detects readiness across all participating GPUs, it responds with release signals to each GPU's synchronizer, triggering concurrent TB dispatch. This mechanism ensures aligned launches across devices and prevents early-issued requests from bypassing potential merge opportunities. 
(2) \textbf{Pre-access synchronization.} Even with synchronized launches, compute divergence may cause TBs to reach memory accesses at different times. When a warp encounters its first \texttt{*.cais} instruction, it sends a synchronization request tagged with the Group ID. Execution proceeds only after all TBs in the group reach the same point. Meanwhile, the warp scheduler can issue independent instructions to hide synchronization latency. 

\reviseD{The synchronization overhead is minimal because it is implemented through the exchange of lightweight empty packets between GPUs and the switch. For each TB, only two empty packets are transmitted between each GPU and the switch. The total latency corresponds merely to the round-trip time between the GPU and the switch, approximately 0.5 µs in our experimental setup, which is negligible compared with the TB execution time. Furthermore, the synchronization scope is strictly confined within each TB group and does not interfere with resource sharing across different TB groups.}

\textbf{TB-Aware Request Throttling.} To avoid stalls from outliers, CAIS introduces a TB-aware request throttling strategy. When a GPU detects that it is ahead of its peer TBs in a mergeable group, it temporarily throttles further requests to allow others to catch up. This feedback is driven by the switch's per-address tracking state and exposed to GPUs via a small control interface. Importantly, this throttling is applied only to mergeable TBs, preserving execution parallelism elsewhere.

\subsubsection{Architecture Support for TB Group Synchronization.} The coordination mechanism is supported by synchronizers on GPUs and a Group Sync Table on the switch, as shown in Fig.~\ref{tbsyncarch}(b). (1)\textbf{On the GPU side,} each device incorporates a synchronizer module that interfaces with the TB and warp schedulers. The synchronizer maintains a small table tracking active TB Groups and handles both pre-launch and pre-access synchronizations by sending TB-group synchronization request to the switch and waiting for a release signal. (2)\textbf{On the switch side}, a lightweight Group Sync Table maintains counters for each active TB Group. When synchronization requests from all GPUs are received for a given Group ID, the switch broadcasts a release signal, allowing execution to proceed. This coordination ensures that mergeable requests from different GPUs arrive within a narrow time window, significantly increasing the success rate of request merging.

\subsection{Graph-Level Dataflow Optimizer}
\label{improvebwutil}

\begin{figure}[!t]
\includegraphics[width=0.5\textwidth]{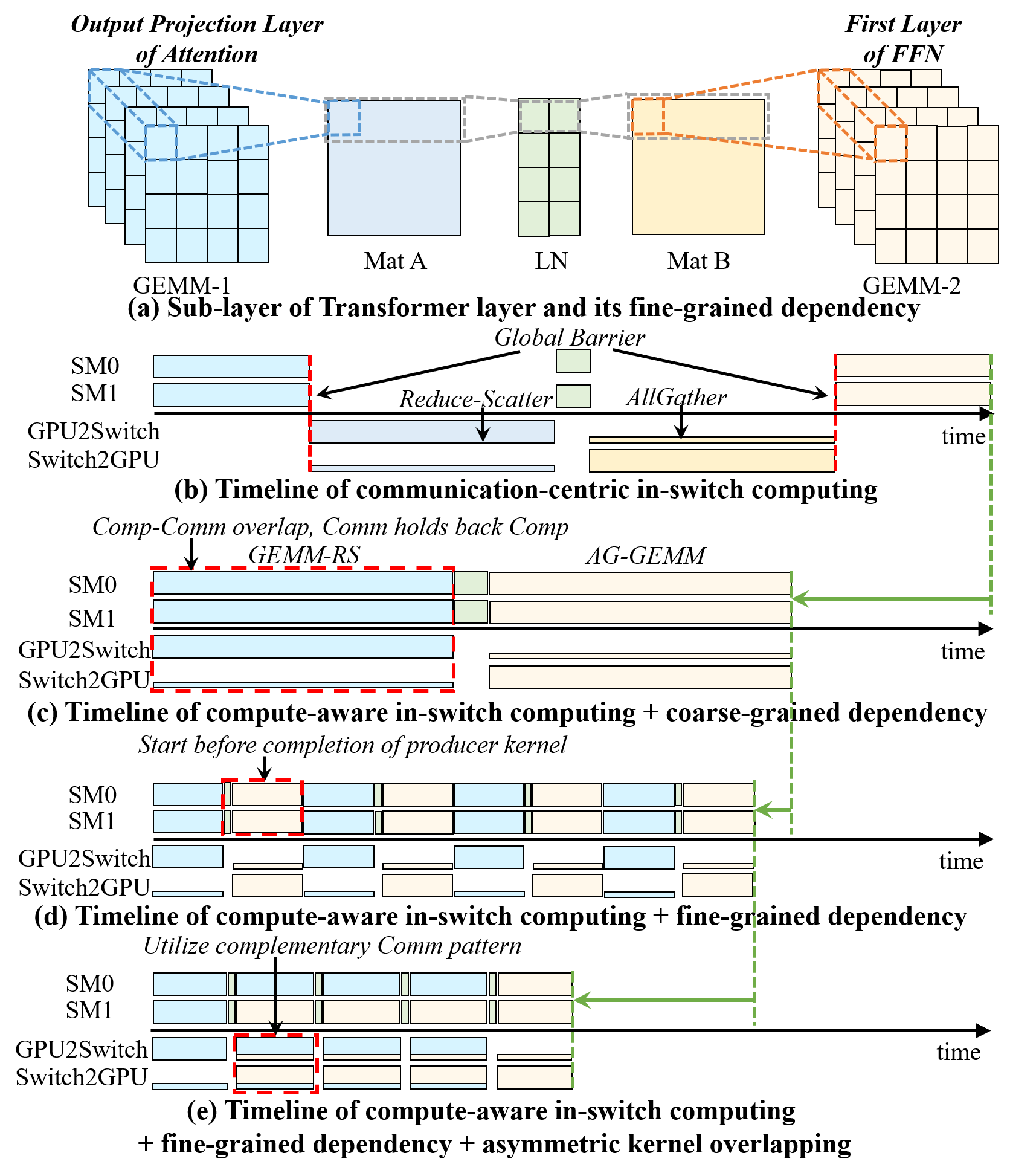}
%\vspace{-.1in}
\caption{Graph-Level Dataflow Optimization. \reviseB{Fine-grained TB-level data dependency enables early launch of consumer TBs before producer kernels complete.}}
\label{fusion}
%\vspace{-.2in}
\end{figure}

\begin{figure}[!t]
\centering
\includegraphics[width=0.5\textwidth]{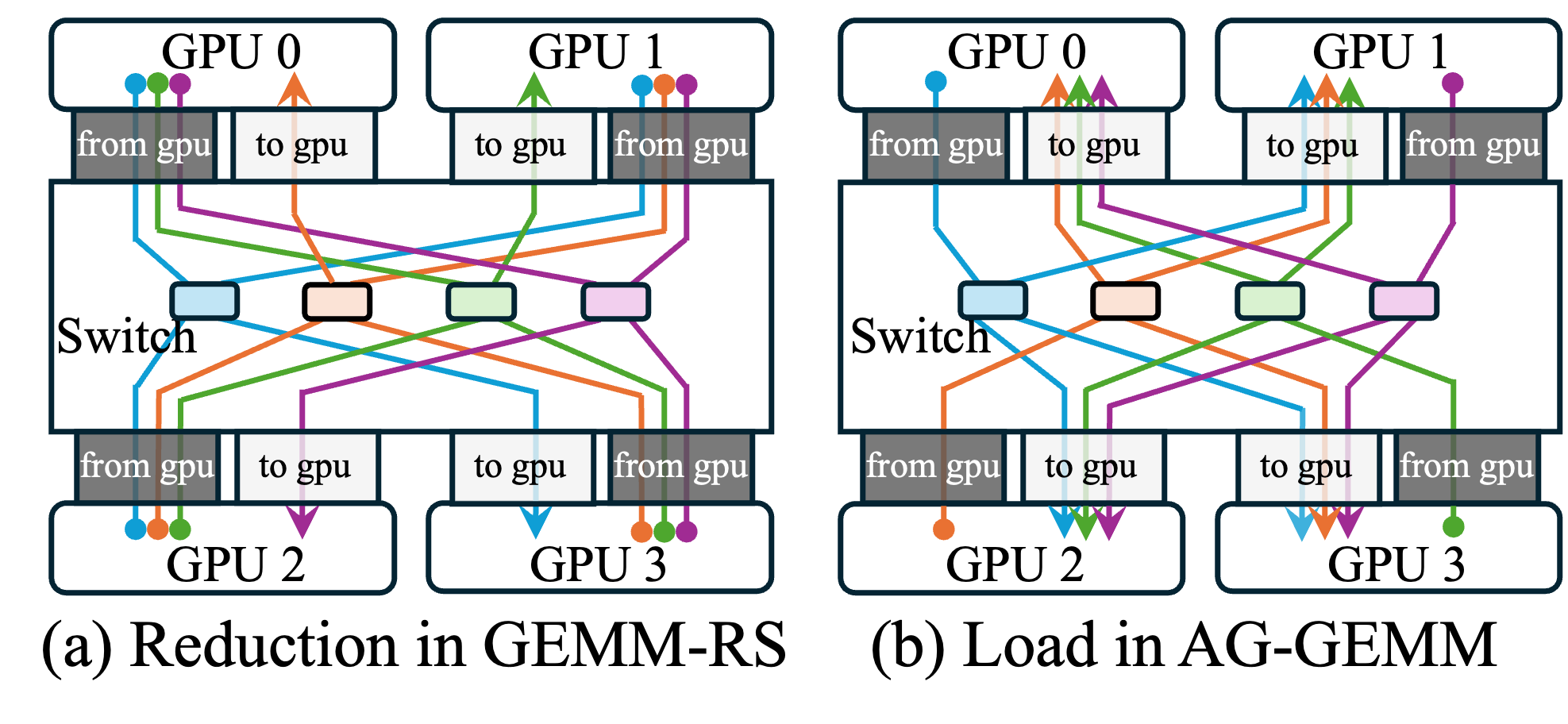}
%\vspace{-.1in}
\caption{Illustration of Asymmetric Traffic.} 
%\vspace{-.25in}
\label{asymmetric}
\end{figure}

CAIS also integrates a graph-level dataflow optimizer to improve the system resource utilization. Graph-level dataflow optimizer supports fine-grained TB-level dependency to unlock tighter kernel fusion opportunities. Built upon the fine-grained TB-level dependency, it introduces Asymmetric Kernel Overlapping to balance complementary traffic between two directions of the inter-chip link, which can significantly improve the overall performance.

\subsubsection{Fine-Grained TB Dependency and Deep Kernel Fusion}

In contrast to coarse-grained kernel-level dependency that require the full completion of a producer kernel before the consumer kernel can start, fine-grained TB-level dependency~\cite{abdolrashidi2021blockmaestro} allows a TB in the consumer kernel ready to be launched as soon as its input data available, without waiting for the entire producer kernel to finish. This capability enables fused execution of multiple dependent kernels.

Fig.~\ref{fusion}(a) illustrates this concept with a portion of a transformer layer, where GEMM-1 computes matrix A, followed by a layer normalization (LN) stage producing matrix B, which is then consumed by GEMM-2. \reviseB{In CAIS, TBs in GEMM-1 collaboratively produce tiles of matrix A; each TB in LN operates on a row of A to generate matrix B; GEMM-2 TBs consume tiles of B to compute the final output matrix. Compared to the coarse-grained execution in Fig.\ref{fusion}(c),} since each TB's input dependencies are localized, execution of GEMM-2 can begin as soon as the corresponding TBs in GEMM-1 and LN complete, \reviseB{unlocking a larger schedule optimization space.} As shown in Fig.\ref{fusion}(d), this fine-grained chaining enables deep kernel fusion and earlier launch of downstream TBs. 

\subsubsection{Asymmetric Kernel Overlapping}
\label{section3C2}

While in-switch merging reduces overall communication volume, it introduces asymmetric bandwidth usage. Operations like GEMM-RS rely on switch-to-GPU reduction traffic, whereas AG-GEMM generates GPU-to-switch load traffic, as is illustrated in Fig.~\ref{asymmetric}. \reviseB{For Fig.~\ref{asymmetric}(a) the reduction operation, operands are read from three GPUs and the result is written back to the destination GPU. This causes the data traffic from GPUs to the switch to be three times higher than that from the switch to the GPUs, creating a bottleneck dominated by the GPU-to-switch path. We refer to this phenomenon as asymmetric traffic.
For Fig.~\ref{asymmetric}(b) the load operation, the situation is exactly the opposite: the switch-to-GPU traffic is three times higher than the GPU-to-switch traffic.}

\reviseB{CAIS exploit the complementary nature of these two traffic patterns to further optimize kernel fusion and improve overall bandwidth utilization.} Using TB-level dependency analysis, CAIS identifies opportunities to pipeline kernels with complementary traffic patterns. For example, when GEMM-RS and AG-GEMM are ready to execute, SMs are partitioned into two groups, each executing one kernel concurrently. This interleaved execution, illustrated in Fig.~\ref{fusion}(e), balances bidirectional link usage: as GEMM-RS emits upstream traffic, AG-GEMM consumes downstream data.

\textbf{Traffic Control.} When kernels with asymmetric communication patterns execute concurrently, contention on the G2S link can still arise, particularly when both load and reduction requests compete for bandwidth. CAIS introduces separate virtual channels for load and reduction traffic and uses round-robin arbitration to avoid head-of-line blocking. 

Together, deep kernel fusion and asymmetric overlapping maximize the bandwidth utilization and compute resources, delivering significant end-to-end performance improvements.

\section{Experimental Methodology}
\label{sec:methodology}

\subsection{Hardware Configuration}
\label{sec:setup}

We simulate an 8-GPU system interconnected via four NVSwitch units, replicating the topology of the NVIDIA DGX-H100~\cite{dgxh100}. To enable accurate modeling, we extend Accel-Sim~\cite{khairy2020accel} with Hopper-specific architectural features and configure the GPU parameters based on the NVIDIA H100 specifications~\cite{h100}. For multi-GPU communication, we integrate Accel-Sim with a customized BookSim2~\cite{jiang2013detailed}, enabling concurrent execution across GPUs connected through a switch-based interconnect. 

\revise{We further modify both Accel-Sim and BookSim2 with custom extensions to support the multimem instructions of NVLS. Specifically, following NVIDIA's NVLS design~\cite{klenk2020network}, we augment the ``router'' in BookSim2 to support in-switch multicast and reduction operations, and extend Accel-Sim to handle the translation from multimem addresses to virtual addresses at the Hub. The quantitative validation of our NVLS simulation is detailed in Section~\ref{sec:validation}.} 
For fair comparison, we also augment T3~\cite{pati2024t3} with NVLS support by adopting the DMA-based NVLS design proposed by NVIDIA~\cite{klenk2020network}.

The NVLink and NVSwitch are modeled using real device parameters. NVLink is configured with a 16B flit size, a single-flit header, and bidirectional data transfer. NVSwitch employs round-robin arbitration with a 40 KB per-port Merge Table (320 entries) and supports routing to forward requests to their target GPUs. Each input port provides eight 256-depth virtual channels. We implement intra-SM request coalescing, aggregating multiple 32B sector requests into packets of up to 128B to emulate NVLink’s burst transfer behavior. Link latency between GPUs and switches (from GPU to switch or from switch to GPU) is configured to 250 ns, resulting in a round-trip latency of approximately 1 µs.

\subsection{Benchmark}
\label{section4B}

\begin{table}[!t]
      \centering
      %\vspace{-.05in}
      \renewcommand{\arraystretch}{1.05}
      \resizebox{0.49\textwidth}{!}{
        \begin{tabular}{|c|c|c|c|c|c|}
\hline
\multicolumn{1}{|c|}{Name} & \multicolumn{1}{c|}{\begin{tabular}[c]{@{}c@{}}Hidden\\ Size\end{tabular}} & \multicolumn{1}{c|}{\begin{tabular}[c]{@{}c@{}}FFN Hidden\\ Size\end{tabular}} & \multicolumn{1}{c|}{\begin{tabular}[c]{@{}c@{}}Attention\\ Heads\end{tabular}} & \multicolumn{1}{c|}{\begin{tabular}[c]{@{}c@{}}Sequence\\ Length\end{tabular}} & \multicolumn{1}{c|}{\begin{tabular}[c]{@{}c@{}}Batch\\ Size\end{tabular}} \\ \hline \hline
Mega-GPT-4B                & 2048                                                                       & 8192                                                                           & 24                                                                             & 1024                                                                           & 16                                                                                                                                                      \\ 
Mega-GPT-8B                & 3072                                                                       & 12288                                                                          & 32                                                                             & 1024                                                                           & 12                                                                                                                                                      \\ 
LLaMA-7B                   & 4096                                                                       & 11264                                                                          & 32                                                                             & 3072                                                                           & 3                                                                                                                                                       \\ \hline
\end{tabular}
        }
        %\vspace{-.08in}
        \caption{LLM Settings Used in Evaluation.}
        \label{llmsetting}
        %\vspace{-.25in}
\end{table}

% \subsubsection{Workload}

We evaluate CAIS using three representative LLMs, summarized in Table~\ref{llmsetting}. Both training and inference phases are evaluated, with inference focusing on the communication-heavy prefill stage. The GEMM kernels are implemented using CUTLASS~\cite{cutlass}. Due to simulator memory constraints and the long simulation time, simulating full-scale state-of-the-art models is infeasible. 
\revise{To address this limitation, we employ scaled-down LLM variants with key matrix dimensions, including hidden size and FFN hidden size, reduced by 50\% compared to state-of-the-art large LLMs. This scaling reduces the computation-to-communication ratio by 50\%. To maintain proportionality, we correspondingly reduce the number of SMs by 50\%. We validate this scaled-down setup in Section~\ref{sec:validation}.}

\begin{figure*}[!t]
\centering
\includegraphics[width=1\textwidth]{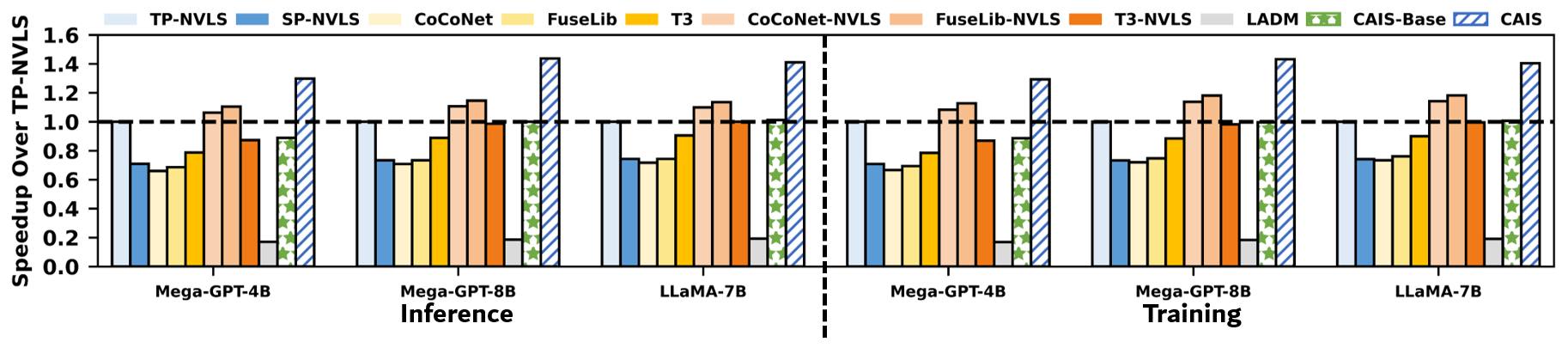}
%\vspace{-.15in}
\caption{\reviseA{End-to-End Model Speedup Across Training and Inference.}}
\label{e2espeedup}
%\vspace{-.1in}
\end{figure*}

\begin{figure*}[!t]
\centering
\includegraphics[width=1\textwidth]{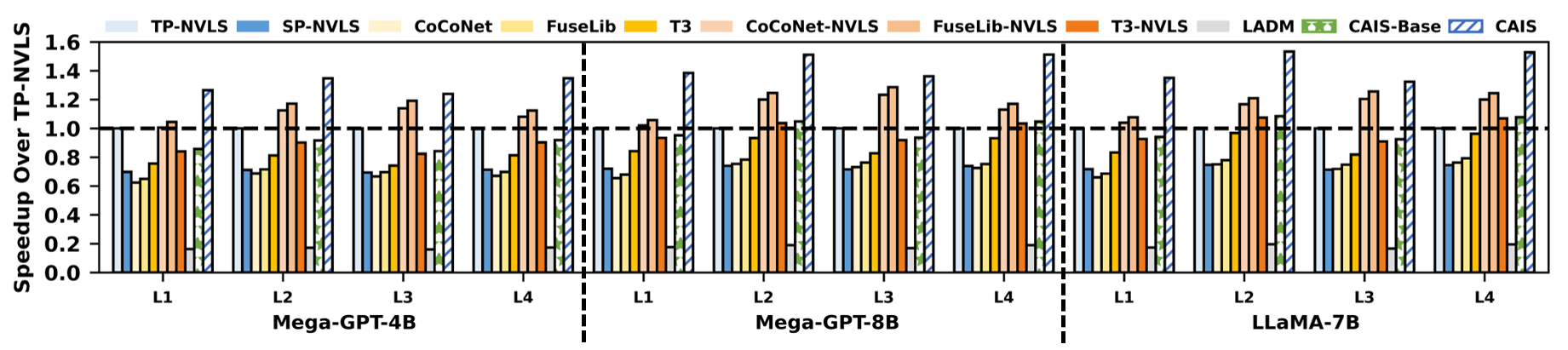}
%\vspace{-.1in}
\caption{\reviseA{Sub-layer Performance Speedup.}}
\label{opspeedup}
%\vspace{-.1in}
\end{figure*}

\subsection{Baseline}
\label{section4C}

\reviseA{CAIS is evaluated against 9 baselines in four categories, including 6 existing works and 3 NVLS-enhanced baselines.} % (e.g., NVLink SHARP) for collective operations: 

\begin{itemize}
\item \emph{Tensor Parallelism with NVLS} includes \textbf{1) Basic TP (TP-NVLS)}~\cite{shoeybi2019megatron} that partitions model layers across GPUs and applies AllReduce to merge intermediate results, and \textbf{2) TP with Sequence Parallelism (SP-NVLS)}~\cite{korthikanti2023reducing} that enhances TP by splitting AllReduce into ReduceScatter and AllGather phases, with layer normalization and dropout/Add operations interleaved to reduce memory footprint. NVLS accelerates these collectives. 

\item \emph{Overlap Solutions} includes \textbf{3) CoCoNet}~\cite{jangda2022breaking} and \reviseC{\textbf{4) FuseLib}}~\cite{punniyamurthy2024optimizing}, both of which enable GEMM-AllReduce overlapping through software scheduling techniques, and \textbf{5) T3}~\cite{pati2024t3} that introduces hardware-assisted fine-grained overlapping between GEMM and ReduceScatter. We extend T3 to also support AG-GEMM overlap in our evaluation. These solutions do not leverage NVLS.

\item \emph{Overlap Solutions with NVLS} includes \textbf{6) CoCoNet-NVLS}, \reviseC{\textbf{7) FuseLib-NVLS}}, and \textbf{8) T3-NVLS}, which are enhanced variants of overlap solutions by integrating NVLS support. \reviseC{As introduced in Sec.~\ref{sec:setup}, CoCoNet-NVLS and FuseLib-NVLS utilize extended multimem instructions, T3-NVLS adopts a DMA-based NVLS design.}

\item \reviseA{\emph{Locality-aware TB schedule} places TBs across GPUs/dies for reducing remote access, where we adopt the SOTA, \textbf{9) LADM}~\cite{khairy2020locality}. LADM cannot utilize NVLS because of its communication-centric design.}

\end{itemize}

\section{Experimental Results}

\subsection{End-to-End and Sub-Layer Speedup}
\label{sec:perf}

\subsubsection{End-to-End Model Speedup}

Figure~\ref{e2espeedup} shows the end-to-end speedup of CAIS over nine baseline methods: TP-NVLS, SP-NVLS, CoCoNet, \reviseC{FuseLib,} T3, CoCoNet-NVLS, \reviseC{FuseLib-NVLS,} T3-NVLS, \reviseA{and LADM}. We also include a stripped-down version, CAIS-Base, which disables the proposed merging-aware TB coordination and graph-level dataflow optimizer. 
For inference, CAIS achieves up to 1.43$\times$, 1.95$\times$, 1.99$\times$, \reviseC{1.92$\times$,} 1.65$\times$, 1.28$\times$, \reviseC{1.24$\times$,} 1.49$\times$, \reviseA{and 7.80$\times$} speedup over these nine baselines, with geometric means of 1.38$\times$, 1.89$\times$, 1.98$\times$, \reviseC{1.90$\times$,} 1.61$\times$, 1.25$\times$, \reviseC{1.21$\times$,} 1.45$\times$, \reviseA{and 7.60$\times$}, respectively. 
For training, CAIS achieves up to 1.44$\times$, 1.96$\times$, 2.03$\times$, \reviseC{1.96$\times$,} 1.65$\times$, 1.30$\times$, \reviseC{1.25$\times$,} 1.49$\times$, \reviseA{and 7.75$\times$} speedup over these baselines with geometric means of 1.37$\times$, 1.89$\times$, 1.96$\times$, \reviseC{1.89$\times$,} 1.60$\times$, 1.23$\times$, \reviseC{1.20$\times$,} 1.45$\times$, \reviseA{and 7.59$\times$,} respectively. 
These results highlight the significant performance advantages obtained by our proposed compute-aware in-switch computing, as well as leveraging architectural and scheduling co-design optimization to improve the temporal alignment and enhance bandwidth utilization.

\subsubsection{Sub-Layer Performance} 

Figure~\ref{opspeedup} reports the performance of four communication-intensive sub-layers across the model execution flow: [L1] Output projection → LayerNorm → First FFN layer (forward); [L2] Second FFN layer → LayerNorm → Input projection (forward); [L3] First FFN layer → LayerNorm → Output projection (backward); [L4] Input projection → LayerNorm → Second FFN layer (backward).

These sub-layers involve GEMM-RS + LN + AG-GEMM, making them ideal candidates for graph-level optimization in CAIS. CAIS consistently outperforms all baselines across these sub-layers, 
with up to 1.53$\times$, 2.05$\times$, 2.11$\times$, \reviseC{2.04$\times$,} 1.67$\times$, 1.36$\times$, \reviseC{1.31$\times$,} 1.51$\times$, \reviseA{and 8.08$\times$} speedup over TP-NVLS, SP-NVLS, CoCoNet, \reviseC{FuseLib,} T3, CoCoNet-NVLS, \reviseC{FuseLib-NVLS,} T3-NVLS, \reviseA{and LADM}, and corresponding geometric means of 1.39$\times$, 1.91$\times$, 1.99$\times$, \reviseC{1.91$\times$,} 1.64$\times$, 1.24$\times$, \reviseC{1.20$\times$,} 1.47$\times$, \reviseA{and 7.90$\times$} speedup.

\subsubsection{Discussions and Analysis}
\label{section5A3}

The improvements over TP-NVLS and SP-NVLS mainly stem from CAIS’s native fine-grained computation-communication overlap derived from the compute-aware in-switch computing. Unlike communication-centric in-switch computing, where the global barrier isolates computation and communication phases, compute-aware in-switch computing only requires the TB-level local barrier, enabling native fine-grained overlapping between computation and communication across different TBs. 
Besides the computation-communication isolation, SP-NVLS also suffers from the low bandwidth utilization incurred from asymmetric communication patterns when accelerating Reduce-Scatter and AllGather with in-switch computing.

CoCoNet-NVLS, \reviseC{FuseLib-NVLS}, and T3-NVLS represent the NVLS-enhanced performance of CoCoNet, \reviseC{FuseLib}, and T3. 
Compared to CoCoNet-NVLS \reviseC{and FuseLib-NVLS}, CAIS supports flexible overlapping that also overlaps the communication with the following GEMM. CAIS also eliminates the need for code modification when implementing kernel fusion and mitigates resource contention between compute and communication kernels in CoCoNet-NVLS. 
\reviseC{Although FuseLib-NVLS executes within a single fused kernel, thereby eliminating kernel-launch overhead and mitigating resource contention, CAIS achieves higher efficiency by enabling more flexible and fine-grained overlap between computation and communication.}
Compared to T3-NVLS, a hardware-based overlapping solution, CAIS demonstrates notable gains. T3-NVLS still suffers from coarse-grained dependency among ReduceScatter, LN, and AllGather stages, which prevents fine-grained optimization opportunities therefore still cannot tackle the asymmetric communication pattern. In contrast, CAIS’s fine-grained TB-level scheduling and asymmetric kernel overlapping unlock additional concurrency and balance inter-GPU bandwidth across two directions. \reviseA{Finally, LADM, though a state-of-the-art locality-aware TB scheduling method for general GPU kernels, focuses on intra-GPU locality rather than inter-GPU communication, and does not leverage NVLS-based acceleration, limiting its applicability to communication-intensive TP workloads.}

Comparing CAIS with CAIS-Base reveals the impact of our merging-aware TB coordination and graph-level dataflow optimizer. The maximum and geomean speedups of 1.49× and 1.45× on end-to-end models 1.46× and 1.43× on end-to-end models inference and 1.46× and 1.42× on training, and 1.51× and 1.47× on sub-layers, respectively. 
\reviseA{These results confirm that merely breaking the global barrier through compute-aware ISA and microarchitectural extensions is insufficient; fully realizing the performance potential requires further optimization within the unlocked scheduling space, leveraging temporal locality and graph-level dataflow integration.}

\subsection{Detailed Performance Analysis}

This section investigates the effectiveness of key architectural techniques within CAIS, including merging-aware TB coordination and graph-level dataflow optimizer.

\subsubsection{Impact of Merging-Aware TB Coordination}
\label{section5B1}

\begin{figure}[!t]
\includegraphics[width=0.5\textwidth]{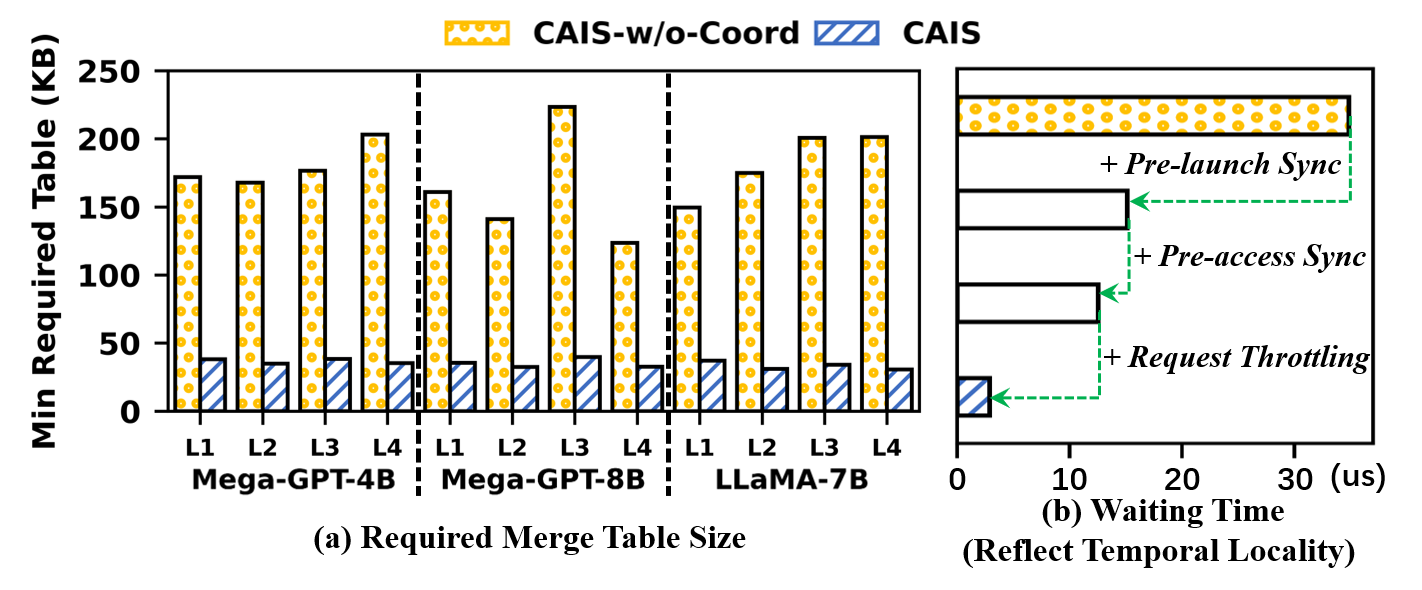}
%\vspace{-.25in}
\caption{(a) Required Merge Table Size with and without Merging-Aware TB Coordination. CAIS reduces the minimal table size needed to merge all eligible requests by 87\%. \reviseC{(b) Ablation studies for TB coordination.}} %, demonstrating high hardware resource efficiency
\label{bufocc}
%\vspace{-.2in}
\end{figure}

\begin{figure}[!t]
\includegraphics[width=0.5\textwidth]{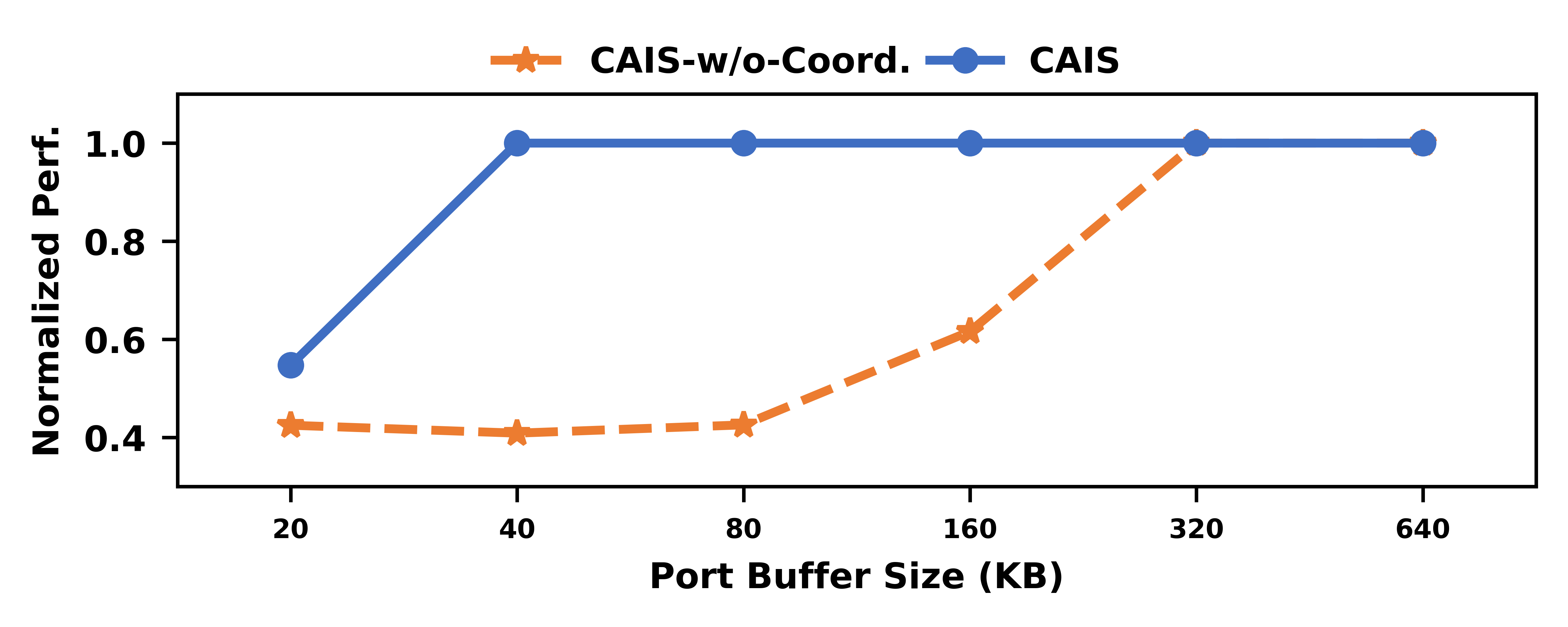}
%\vspace{-.25in}
\caption{Performance Sensitivity to Merge Table Size. CAIS maintains high performance with small table sizes, while the uncoordinated version degrades rapidly.} %, highlighting the effectiveness of temporal alignment
\label{bufperf}
%\vspace{-.2in}
\end{figure}

The merging-aware TB coordination mechanism significantly reduces the waiting time for request merging at the switch by improving the temporal alignment of memory requests across GPUs. Figure~\ref{bufocc} reports the minimal Merge Table size required to merge all mergeable requests for each sub-layer. Without coordination (CAIS-w/o-Coord), the minimal required table size can reach up to 250 KB per port. With coordination enabled, the minimal required table size drops below 40 KB across all ports, which is an 87\% reduction in minimal required table size. This result suggests that our coordination strategy can achieve a more effective use of limited switch resources. Figure~\ref{bufocc}(a) also demonstrates that the minimal required table sizes of CAIS are insensitive to the model sizes and configurations, and are consistently below 40 KB under different model sizes and configurations.

\reviseC{Figure~\ref{bufocc}(b) further evaluates the effectiveness of each optimization. We measure the improvement using the average waiting time, defined as the delay between the earliest and latest requests targeting the same address. This metric directly reflects the temporal locality optimized by TB coordination. The results show that each optimization step progressively enhances temporal locality, reducing the waiting time from 35 µs to less than 3 µs.}

Figure~\ref{bufperf} complements this analysis by showing how coordination affects performance under varying Merge Table sizes for the LLaMA-7B model. Merging-aware TB coordination maintains high performance even when the switch buffer is small, while the uncoordinated version degrades rapidly. These comparisons emphasize the importance of merging-aware TB coordination for compute-aware in-switch computing.

\subsubsection{Impact of Graph-Level Dataflow Optimizer}

Our proposed graph-level dataflow optimizer allows concurrent execution of dependent kernels with complementary asymmetric communication patterns. This optimization improves overall bandwidth utilization by balancing traffic across GPU-to-switch and switch-to-GPU links.

\begin{figure}[!t]
\centering
\includegraphics[width=0.5\textwidth]{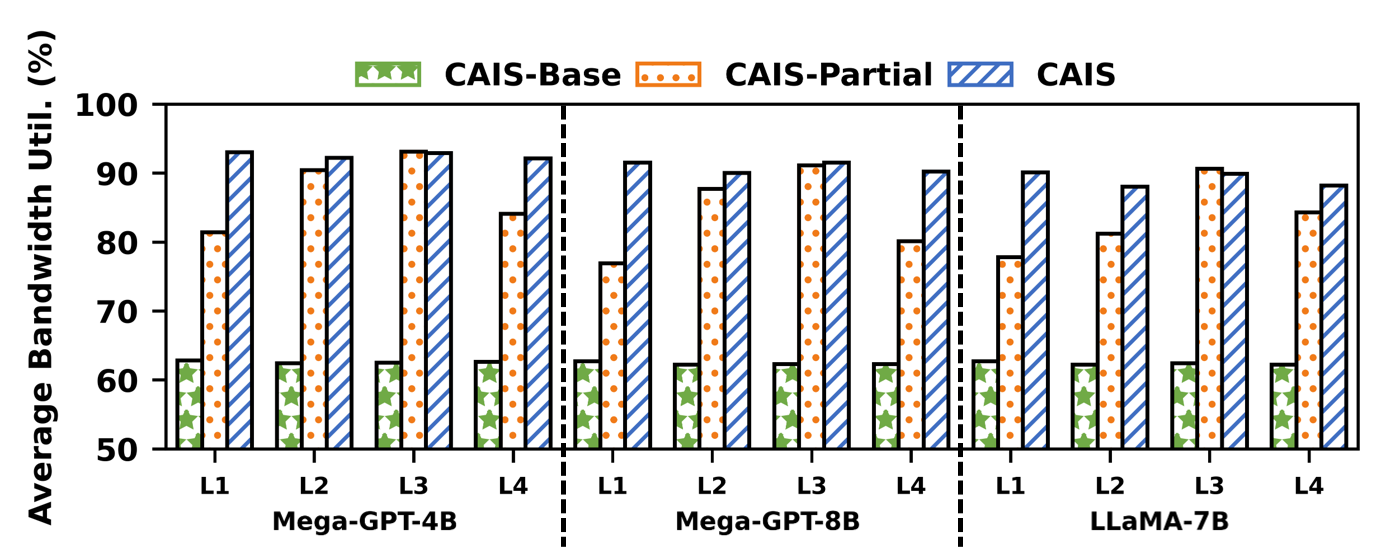}
%\vspace{-.12in}
\caption{Average Bandwidth Utilization per Sub-layer.}
\label{bwutil}
%\vspace{-.2in}
\end{figure}

Figure~\ref{bwutil} illustrates this effect by comparing the average bandwidth utilization, which is the average across all links and two directions for each link, for all sub-layers of three configurations: (a) CAIS-Base, (b) CAIS with graph-level dataflow optimizer but without traffic control (CAIS-Partial), and (c) full CAIS. Bandwidth utilization improves from 62.4\% (CAIS-Base) to 84.7\% (CAIS-Partial) and 90.2\% (CAIS). The gain from CAIS-Base to CAIS-Partial comes from asymmetric kernel overlapping that tackles the imbalance data movement in both link directions, while the final jump to CAIS reflects the benefit of traffic control.

\begin{figure}[!t]
\centering
\includegraphics[width=0.5\textwidth]{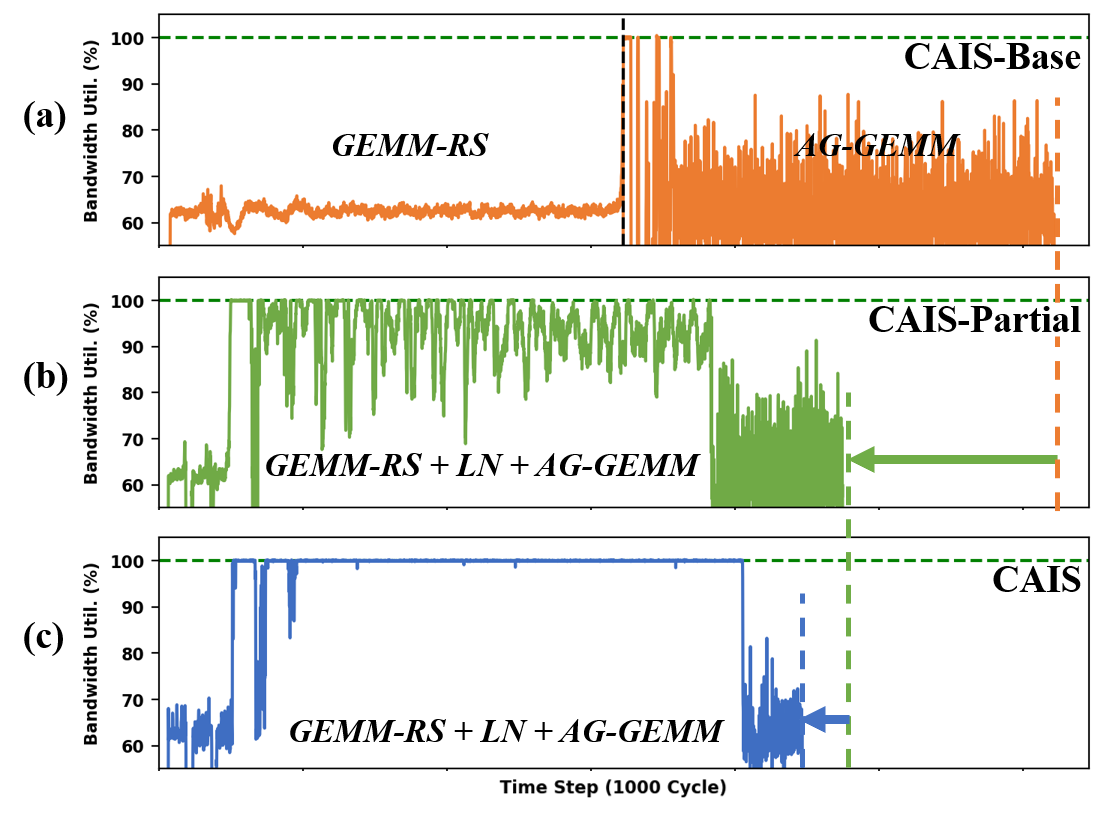}
%\vspace{-.12in}
\caption{Bandwidth Utilization over Time for (a) CAIS-Base, (b) CAIS-Partial, and (c) CAIS.}
\label{bwtrace}
%\vspace{-.2in}
\end{figure}

To further analyze the sustained behavior of these improvements, Figure~\ref{bwtrace} presents the bandwidth utilization over time for the L2 sub-layer of LLaMA-7B. CAIS maintains near-peak utilization ($\sim$100\%) during steady-state operation, while the partial configuration (CAIS-Partial) suffers dips due to contention. The base configuration shows the lowest and most fluctuating utilization. This demonstrates the importance of dataflow optimization and traffic control.

Together, these analyses demonstrate that the graph-level dataflow optimizer is essential to unlocking the full potential of compute-aware in-switch computing.

\subsection{Scalability Analysis}

\subsubsection{Performance Scalability} 
\begin{figure}[!t]
  \centering
  %\vspace{-.2in}
  \begin{minipage}{0.23\textwidth}
    \centering
    \includegraphics[width=\textwidth]{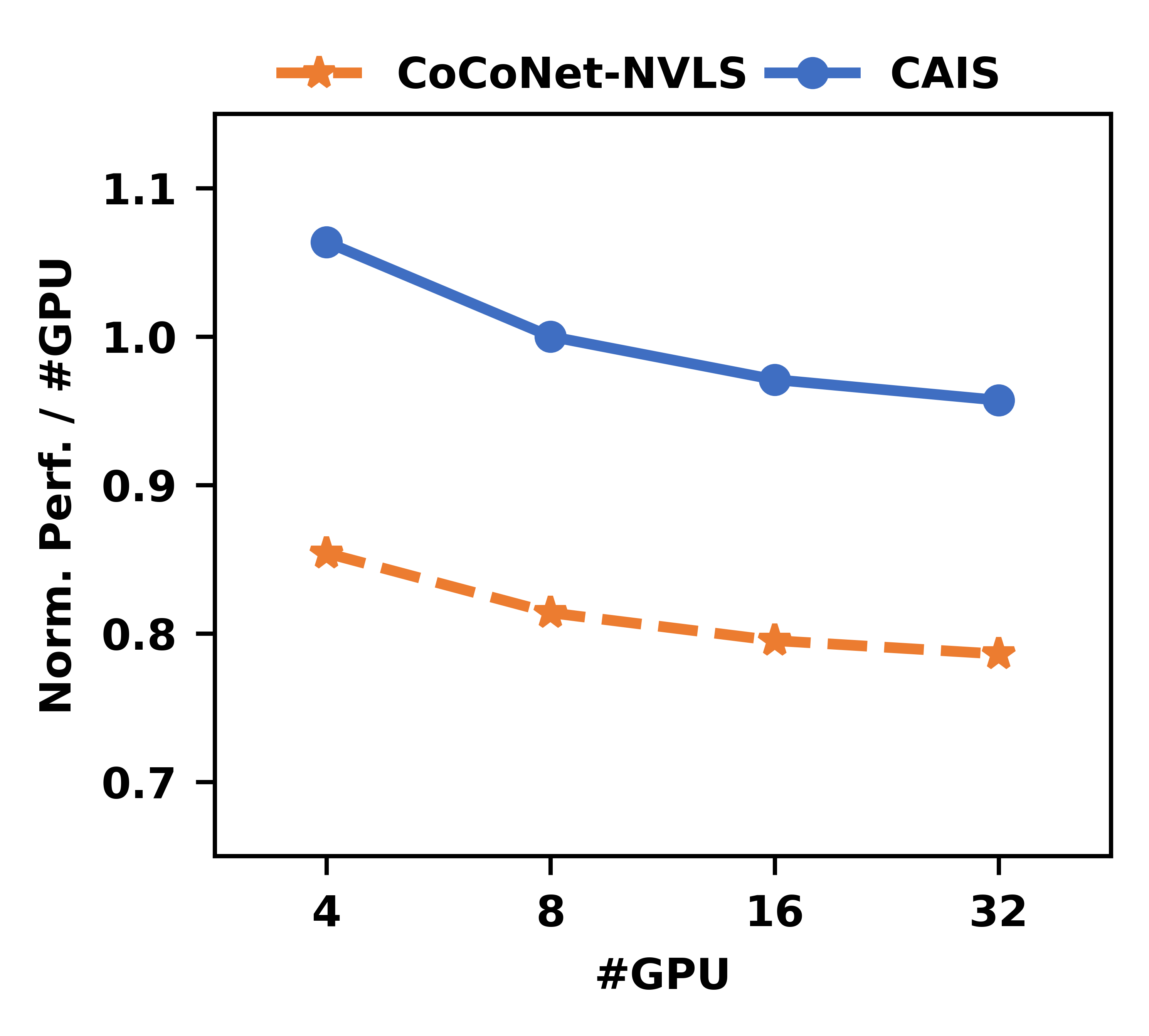}
    %\vspace{-.22in}
    \caption{Scalability with Increasing GPU Count}
    \label{scalability}
  \end{minipage}
  %\vspace{-.12in}
  \hfill
  \begin{minipage}{0.24\textwidth}
    \centering
    \includegraphics[width=\textwidth]{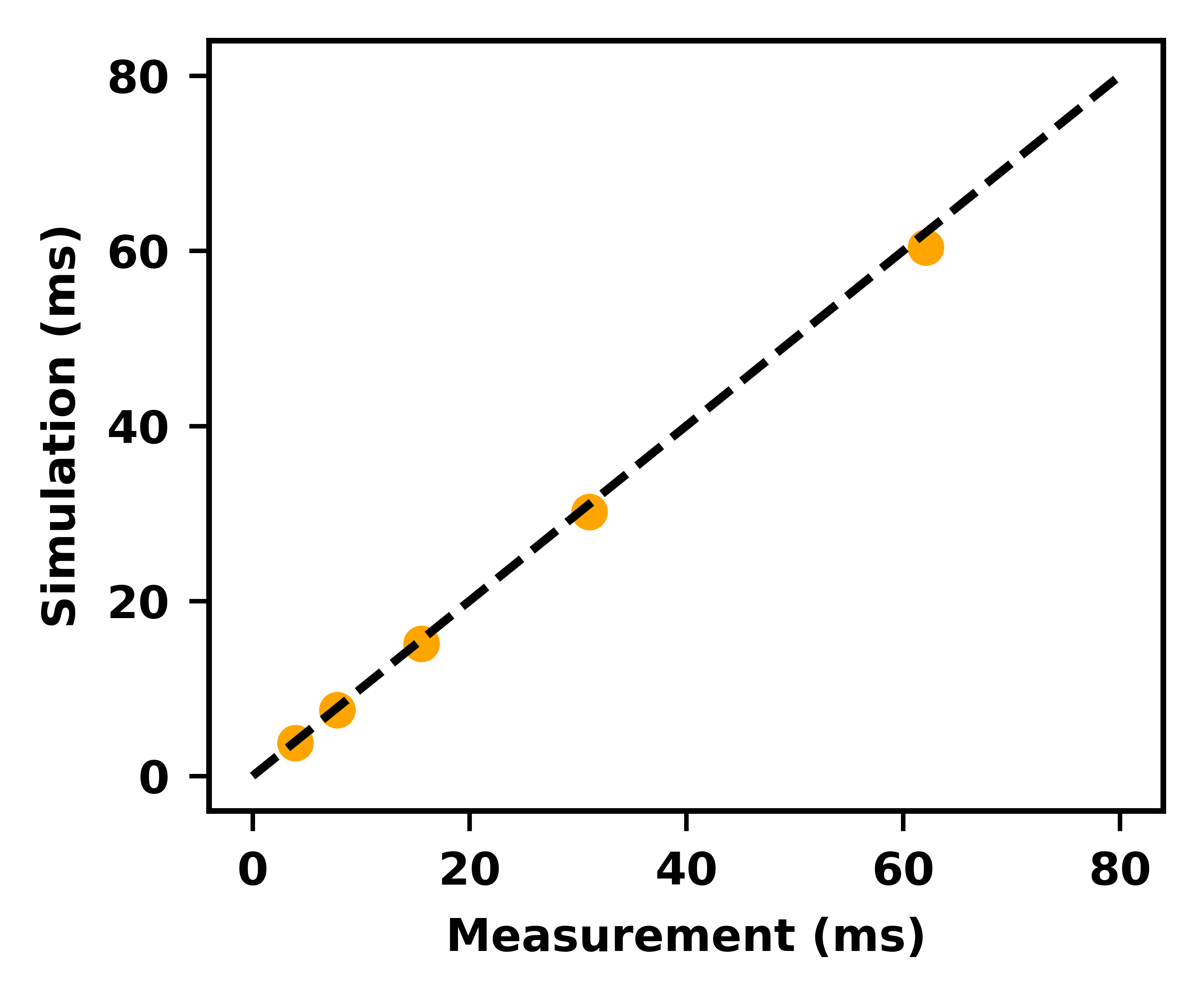}
    %\vspace{-.2in}
    \caption{\revise{Validation of Our Simulated NVLS.}}
    \label{measurement}
  \end{minipage}
  %\vspace{-.12in}
\end{figure}

To demonstrate the performance scalability of CAIS, we evaluate CAIS and CoCoNet-NVLS across different numbers of GPUs based on the LLaMA-7B model. We also scale the model's hidden dimension proportionally to the number of GPUs to avoid under-utilization of computation resources. Figure~\ref{scalability} shows the performance scalability of CAIS and CoCoNet-NVLS, where we measure per-GPU computation throughput normalized to 8-GPU CAIS. It shows that the per-GPU throughput decreases slightly when increasing the number of GPUs. Even with 32 GPUs, the performance drop is still within 5\% compared with 8 GPUs. Both CAIS and CoCoNet-NVLS can consistently have superior performance, regardless of the number of GPUs.

\subsubsection{Hardware Cost Scalability} CAIS hardware exhibit excellent scalability as the number of GPUs increases, owing to a constant and low upper bound on the total required merging table size at the system level. This table size corresponds to the amount of data associated with all outstanding remote requests in the system. CAIS leverages merge-aware coordination to synchronize potentially mergeable requests before they are issued to the switch. This ensures that all GPUs issue outstanding requests for the same set of data, enabling deterministic merging at the switch. As a result, the total required merging table size is bounded by the outstanding remote requests from a single GPU, rather than scaling with the number of GPUs. In addition, CAIS’s request throttling mechanism further limits the number of outstanding remote requests per GPU, reducing the actual memory footprint. In our evaluated 8-GPU system, the system-wide upper bound is 1280 KB, corresponding to just 40 KB per switch port. Importantly, this bound remains unchanged even as the number of GPUs increases. Therefore, as the system scales, the relative hardware overhead decreases, since the size of the base switch circuitry grows with GPU count while the enhanced logic remains constant. This bounded memory overhead ensures the scalability and practicality of CAIS’s hardware extensions across large-scale multi-GPU deployments.

\subsection{Hardware Overhead}

We evaluate our hardware overhead under TSMC’s 12nm process technology. Our hardware modification for the switch occupies about 0.50$mm^2$, which is less than 1\% of NVIDIA’s NVSwitch die~\cite{hcs2018nvswitch,klenk2020network}. On the GPU side, the added logic for TB-group-based synchronization consumes only $0.019mm^2$ per die, less than 0.01\% of the H100 GPU area. These results confirm that our proposed architectural support is cost-effective and hardware-feasible.

\subsection{\revise{Methodology Validation}}
\label{sec:validation}

This section validates our experimental methodology from two aspects: 1) fidelity of a scaled-down configuration for both LLM size and GPU resources, and 2) accuracy of our NVLS-enabled simulator with support for \texttt{multimem} instructions.

\revise{For scaled-down setup, we compare two systems: a full-scale GPU executing a full-sized LLM that is feasible on Accel-Sim, and a half-scale system with 50\% fewer SMs running the same model with matrix dimensions halved. As reported in Table~\ref{scaledown}, the half-scale configuration faithfully reproduces full-scale speedup ordering and magnitudes, preserving system-level behavior and key insights derived.}

\revise{To validate NVLS support in simulation, we measure All-Reduce performance using NCCL~\cite{nccl} on both real hardware and our simulator across message sizes from 1~GB to 16~GB (1, 2, 4, 8, 16~GB). As shown in Fig.~\ref{measurement}, simulated results closely match the real-system measurements, yielding high fidelity with an average error of only 3.87\%.}

\begin{table}[]%\color{blue}
\centering
\renewcommand{\arraystretch}{1.05}
      \resizebox{0.47\textwidth}{!}{
\begin{tabular}{|c||c|c|c|c||c|}
\hline
\multicolumn{1}{|l||}{\revise{Setup}} & \begin{tabular}[c]{@{}c@{}}\revise{Hidden}\\ \revise{Size}\end{tabular} & \begin{tabular}[c]{@{}c@{}}\revise{FFN Hidden}\\ \revise{Size}\end{tabular} & \begin{tabular}[c]{@{}c@{}}\revise{Attention}\\ \revise{Heads}\end{tabular} & \multicolumn{1}{l||}{\revise{\# SM}} & \begin{tabular}[c]{@{}c@{}}\revise{CAIS Speedup}\\ \revise{Over TP-NVLS}\end{tabular} \\ \hline \hline
\revise{Full} & \revise{8192} & \revise{22528} & \revise{64} & \revise{132} & \revise{\textbf{1.43}} \\ \hline
\revise{Half} & \revise{4096} & \revise{11264} & \revise{32} & \revise{66} & \revise{\textbf{1.40}}                                                                               \\ \hline
\end{tabular}}
%\vspace{-.07in}
\caption{\revise{Experimental Validation of Scaling-down Setup.}}
%\vspace{-.2in}
\label{scaledown}
\end{table}

% \begin{figure}[!t]
% \centering
% \includegraphics[width=0.25\textwidth]{_fig/nvls_validatioin.png}
% \vspace{-.1in}
% \caption{\revise{Validation of Our Simulated NVLS.}}
% \label{measurement}
% \vspace{-.2in}
% \end{figure}

\section{Related Work}
\label{section6}

To mitigate the communication bottleneck, prior works have proposed various optimizations. These can be broadly categorized into two types: 1) communication-computation overlapping~\cite{jangda2022breaking,chen2024centauri,pati2024t3,wang2022overlap,punniyamurthy2024optimizing} and 2) memory efficiency improvement~\cite{korthikanti2023reducing,li2021sequence}. CoCoNet~\cite{jangda2022breaking} pioneered the idea of overlapping computation and AllReduce in TP through software pipelining, \reviseC{while \cite{punniyamurthy2024optimizing} further reduces kernel-launch overhead using a similar software-based approach.} T3~\cite{pati2024t3} proposes hardware primitives to enable fine-grained kernel overlap. Centauri~\cite{chen2024centauri} proposes overlapping in hybrid parallelism. 
However, directly applying these techniques to NVLS still fails to achieve effective overlap, since NVLS follows a communication-centric philosophy that lacks awareness of computation semantics, a limitation that CAIS explicitly resolves. 
\reviseD{Although ACE~\cite{rashidi2021enabling} adopts an in-network computing approach, it primarily reduces DRAM data accesses.} 
\reviseA{Some prior studies have also explored locality-aware TB scheduling~\cite{arunkumar2017mcm,milic2017beyond,khairy2020locality}, but they primarily reduce remote memory access volume across GPUs. CAIS aims to enhance the temporal locality for better request merging.}

\section{Conclusion}

We present CAIS, a compute-aware in-switch computing framework that provides aligned communication mode with LLM computation kernels. CAIS advances NVLS from a communication-centric primitive accelerator to a compute-integrated, semantics-aligned in-switch computing framework, enabling fine-grained overlap that improves end-to-end LLM system efficiency.
CAIS introduces the compute-aware ISA and microarchitecture extension to enable compute-aware in-switch computing, and integrates merge-aware TB coordination and graph-level dataflow optimizer to unlock the full potential of compute-aware in-switch computing. Our evaluation on representative LLMs shows that CAIS delivers significant performance gains over the SOTA solutions.

\section{ACKNOWLEDGMENT}

We sincerely thank the anonymous HPCA’26 reviewers
for their valuable suggestions that improved the paper. This work is supported by the Shanghai QiYuan Innovation Foundation and Natural Science Foundation of Shanghai (NSFS) grant 25ZR1402275, National Natural Science Foundation of China (NSFC) grant (62502305 and 62222210), National Key R\&D Program of China under Grant 2022YFB4501400.

%%%%%%% -- PAPER CONTENT ENDS -- %%%%%%%%

\bibliographystyle{IEEEtranS}
\bibliography{refs}

\end{document}